\documentclass[prb,reprint,notitlepage,amsmath,amssymb,amsfonts,%
superscriptaddress,nofootinbib]{revtex4-2}

\usepackage{color}
\usepackage{graphicx}   % Enables the inclusion of graphics
\usepackage{array}
\usepackage{dcolumn}    % alligning table entries w.r.t decimal point
\usepackage{bm}         % bold math
\usepackage[normalem]{ulem}

\usepackage[bookmarks=false,hidelinks,unicode=true]{hyperref}% add hypertext capabilities
\hypersetup{colorlinks=true,allcolors=blue} 

\newcommand{\blue}[1]{{#1}}
\newcommand{\green}[1]{{#1}}

\allowdisplaybreaks     % For allowing pagebreaks in multiline equations

\begin{document}
\title{Multiresonant metasurfaces for arbitrarily-broadband pulse chirping and dispersion compensation}

%\title{Achromatic Metasurfaces for Broadband Steering and Focusing: \\ Squeezing a Wedge and Lens into a Surface}

\author{Odysseas Tsilipakos}\email{otsilipakos@eie.gr}
\affiliation{Theoretical and Physical Chemistry Institute, National Hellenic Research Foundation, GR-11635 Athens, Greece}
\affiliation{Institute of Electronic Structure and Laser, Foundation for Research and Technology Hellas, GR-70013 Heraklion, Greece}

\author{Thomas Koschny}
\affiliation{Ames Laboratory---U.S. DOE and Department of Physics and Astronomy, Iowa State University, Ames, Iowa 50011, USA}

\date{\today}% It is always \today, today,

\begin{abstract}
We show that ultrathin metasurfaces with a specific multiresonant response can enable simultaneously arbitrarily-strong and arbitrarily-broadband dispersion compensation, pulse (de-)chirping and compression or broadening. This breakthrough overcomes the fundamental limitations of both conventional  non-resonant approaches (bulky) and modern singly-resonant metasurfaces (narrowband) for quadratic phase manipulations of electromagnetic signals. The required non-uniform trains of resonances in the electric and magnetic sheet conductivities that completely control phase delay, group delay, and chirp, are rigorously derived and the limitations imposed by fundamental physical constraints are thoroughly discussed. Subsequently, a practical, truncated approximation by finite sequences of physically-realizable linear resonances is constructed and the associated error is quantified. By appropriate spectral ordering of the resonances, operation can be achieved either in transmission or reflection mode, enabling full space coverage. The proposed concept is not limited to dispersion compensation, but introduces a generic and powerful ultrathin platform for the spatio-temporal control of broadband real-world signals with a myriad of applications in modern optics, microwave photonics, radar and communication systems.
\end{abstract}
%leading to unprecedentedly thin realizations

\maketitle

\section{Introduction\label{sec:intro}}
Metasurfaces (MSs), ultrathin artificial media composed of subwavelength resonant meta-atoms, are being extensively studied for a myriad of applications \cite{Glybovski:2016,Chen:2016,He2019research,Sun2019aop,Tsilipakos:2020progress}.
Despite their ultrathin nature, MSs can impart a nontrivial phase delay on the impinging wave due to the meta-atom resonance, which when spatially modulated is typically exploited for wavefront manipulation \cite{Estakhri:2017ACS,Asadchy:2017ACS}. However, this resonant phase delay is inherently dispersive, resulting in narrowband operation. Therefore, conventional metasurfaces can sustain their functionality over very limited bandwidths and fail to perform well for real-world signals which necessarily have significant temporal bandwidth. Thus, researchers have recently focused on the search for broadband (achromatic) MSs that are suitable for practical applications.

\begin{figure}
\centering
\includegraphics[width=8.5cm]{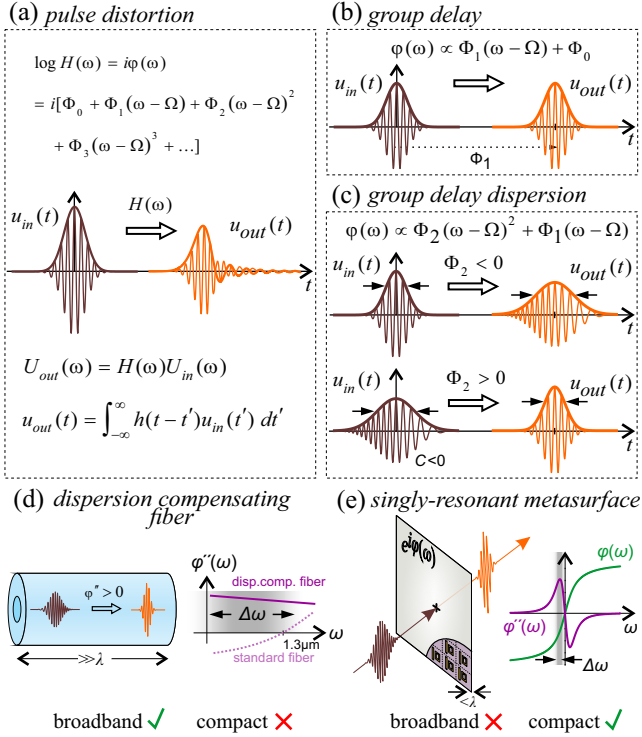}
\caption{\label{fig:Concept} Motivation and scope of current work.
(a)~Broadband pulse (centered at frequency $\Omega$) interacting with a medium described by a transfer function of the form $H(\omega)=\exp[i\phi(\omega)]$, where $\phi(\omega)=\Phi_0+\Phi_1(\omega-\Omega)+\Phi_2(\omega-\Omega)^2+\ldots\ $ includes arbitrarily-high orders resulting in pulse distortion. Impact of leading terms of the Taylor expansion on the output pulse shape  $u_\mathrm{out}(t)$. (b)~The first-order term describes pulse delay by $\phi'(\Omega):=d\phi/d\omega\!\!\!\mid_{\Omega}\,\,=\Phi_1$. The constant $\Phi_0$ leads to a simple shift of the carrier oscillation under the envelope. (c)~The second-order term ($\Phi_2$, group delay dispersion) describes pulse chirping (variation of instantaneous frequency along pulse), typically leading to pulse broadening due to the different delay of constituent frequency components (useful for e.g. chirped pulse amplification). For a pre-chirped pulse with chirp parameter $C$, pulse compression can be achieved when $\Phi_2 C<0$ (useful for e.g. dispersion compensation). (d,e)~Prototypical examples of physical systems for dispersion compensation. (d)~Dispersion compensation fiber: The response is broadband but the system is bulky. (e)~Conventional singly-resonant metasurface: Thin structure but narrowband operation. Electromagnetically-induced-transparency (EIT) \cite{Dastmalchi:2014} or Huygens metasurfaces \cite{Decker:2015} can help to capture the peak of group delay dispersion (GDD) under high transmission.}
\end{figure}

Prominent examples of broadband functionalities reported thus far with MSs include wavefront manipulation (e.g., beam steering/splitting, focusing and imaging) \cite{Wang:2017,Chen:2018,Shrestha:2018,Fathnan:2020aom,Tsilipakos:2020} and pulse delay \cite{Tsilipakos:2021}. Both require a spectrally-constant group delay by the MS [or, equivalently, a linear phase profile $\phi(\omega)$], in order to uniformly delay all frequency components of the broadband input pulse and avoid pulse distortion [Fig.~\ref{fig:Concept}(a),(b)]. However, a wider class of very important applications depend on a quadratic phase profile, \blue{e.g., dispersion compensation, chirped pulse amplification (CPA), and in general any application requiring control over the instantaneous frequency (chirp) and temporal duration of a broadband input pulse through pulse chirping/de-chirping} [Fig.~\ref{fig:Concept}(c)]. Such operations conventionally require lengthy bulk media, e.g., dispersion compensation fibers in optical telecommunications [Fig.~\ref{fig:Concept}(d)].
%microwave photonics and radar systems.

Thus far, the approaches to dispersion compensation with metasurfaces are scarce \cite{Dastmalchi:2014,Decker:2015,Divitt:2019,Rahimi:2016}. They are either very narrowband or do not guarantee pulse integrity. This is because the phase profile is not designed to be purely quadratic across a wide bandwidth (accompanied by a flat amplitude response); rather, typically a single frequency featuring maximum group delay dispersion (GDD) is being exploited  \cite{Decker:2015}, Fig.~\ref{fig:Concept}(e). In Ref.~\onlinecite{Divitt:2019} a broadband pulse is separated into many frequency components and each of them is handled separately by a different, narrowband sub-metasurface. Note that electromagnetically-induced-transparency (EIT) \cite{Dastmalchi:2014} or Huygens' metasurfaces \cite{Decker:2015} can help to capture the peak of GDD under high transmission. Operation in reflection is not being discussed.
%In addition, these approaches are are limited to operation in transmission

In this work, we present a solution to these problems. We show that by using multiresonant metasurfaces we can overcome the limitations of both traditional, non-resonant approaches (bulky) and modern, singly-resonant metasurfaces (narrowband). Our approach allows to design MSs that implement a general quadratic phase profile which is both arbitrarily strong (despite the ultrathin nature) and (almost) arbitrarily broadband, controlled at will by the number and spacing of the implemented resonances. We derive an explicit construction for the sheet conductivities of a multiresonant surface that can completely control the first three dispersion parameters (phase delay, group delay, and chirp) and discuss the fundamental limitations of physically-possible phase manipulations of broadband chirped pulses by such metasurfaces. We subsequently approximate by finite sequences of physically-realizable Lorentzian resonances and rigorously quantify the associated error. Both signs of GDD can be implemented and both operation in transmission and reflection mode; as a result, full-space coverage can be provided. Importantly, the required phase delay is solely provided by the resonances implemented on the surface itself. Thus, the proposed surfaces are essentially 2D, apart from a small finite thickness to allow for implementing magnetic polarizability without magnetic materials.

\blue{Note that using multiple Lorentzian resonances is the basis of many models meant to capture the response function of solids (susceptibility or permittivity) as accurately as possible. For instance, the Brendel-Borrman model takes into account statistical variations in the vibrational frequencies of amorphous media and models the resulting inhomogeneous broadening by convolving the different Lorentzians with a Gaussian function centered at the respective resonant frequency \cite{Brendel:1992}. Inhomogeneous broadening should have implications for our work as well, since in a realistic metasurface deviations in the meta-atom dimensions along the metasurface would lead to linewidth broadening. In the process of deriving such models, it is important to adhere to the restrictions of causality and the Kramers-Kronig criteria \cite{Orosco:2018}. This means symmetrizing the spectrum of the response function and removing any singularity in the upper complex half-plane \cite{Orosco:2018}, which are common elements with our work. Furthermore, ending up with a causal and real-valued time-domain representation of the material response function is also important in the context of time-domain computational electromagnetics (e.g. the Finite-Difference Time Domain Method). In such cases, the efficient incorporation of the material model in the numerical algorithm becomes important as well \cite{Prokopeva:2022}.}

%\section{Results}
\section{Theory of Multiresonant Metasurfaces for a Quadratic Phase Profile}
\begin{figure*}
\centering
\includegraphics[width=\textwidth]{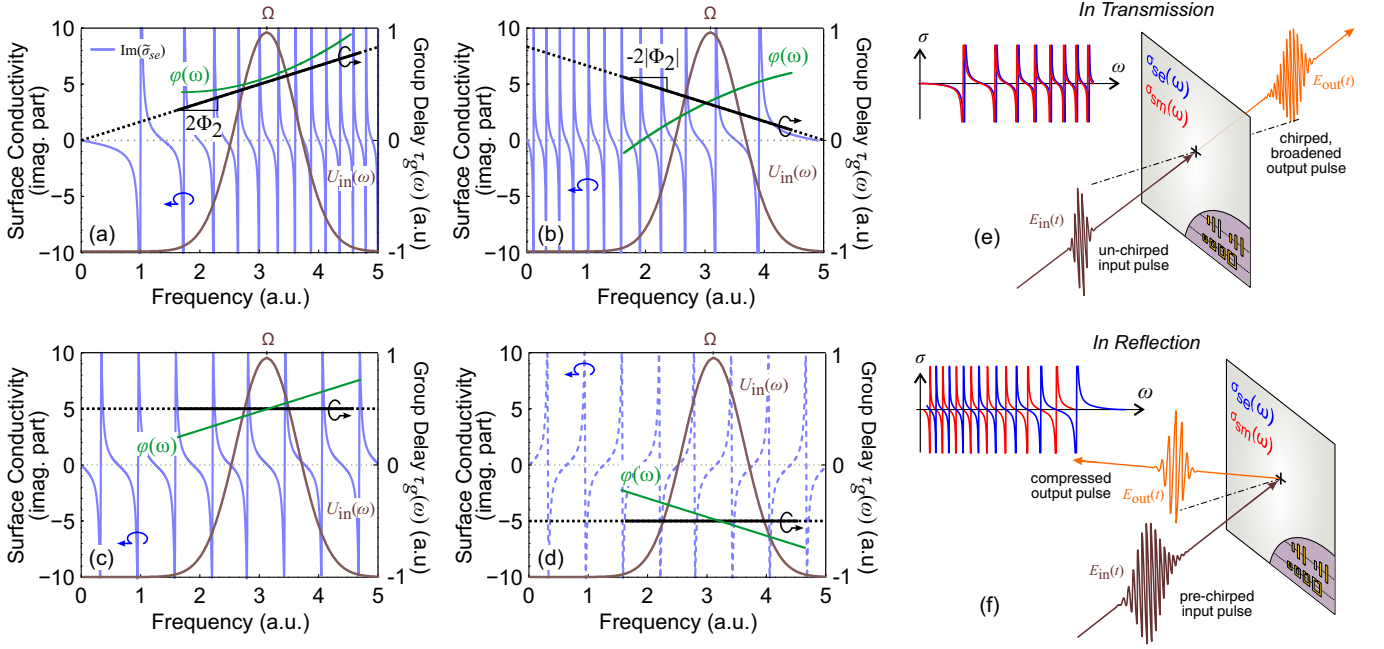}
\caption{\label{fig:Functions} Proposed multiresonant approach for positive/negative chirp and reflection/transmission operation. Generic multiresonant metasurface with a quadratic transmission/reflection phase profile [$\phi(\omega)=\Phi_2(\omega-\Omega)^2+\Phi_1(\omega-\Omega)+\Phi_0$] providing a spectrally-linear group delay of slope $\mathrm{GDD}=2\Phi_2$ for pulse (de-)chirping and pulse broadening/compression. (a)~Surface conductivity with multiple resonances of decreasing spacing and linewidth for a positive (constant) slope of the group delay. (b)~Multiple resonances of increasing spacing and linewidth for a negative slope.
(c)~A constant group delay is achieved with equally-spaced resonances \cite{Tsilipakos:2018}. (d)~A negative constant group delay would require anti-resonances.
(e)~An initially un-chirped pulse is temporally broadened and chirped after interacting with the metasurface. Operation in transmission requires overlapping (matched) electric and magnetic resonances. (f)~A pre-chirped pulse can be compressed when $\Phi_2$ is of opposite sign. This is the basis of dispersion compensation in e.g. optical fiber systems. Operation in reflection requires interleaved (antimatched) resonances.}
\end{figure*}

The main elements of our approach are presented in Fig.~\ref{fig:Functions}. Implementing a surface with a very specific multiresonant surface conductivity can provide a perfectly quadratic phase profile $\phi(\omega)=\Phi_2(\omega-\Omega)^2+\Phi_1(\omega-\Omega)+\Phi_0$ [Fig.~\ref{fig:Functions}(a),(b)]. The corresponding slope (GDD) is constant and equals $2\Phi_2$. By making the resonant features denser(sparser) as the frequency increases, a positive(negative) chirp can be implemented; the linewidths of the resonances follow a similar trend.  Note that the simpler, special case of equally-spaced resonances would result in a constant (positive) group delay that can be used for delaying broadband pulses \cite{Tsilipakos:2018,Ginis:2016}, see Fig.~\ref{fig:Functions}(c). In addition, the linewidth (imaginary part of the complex frequency) is constant for all resonances. A negative constant group delay would require anti-resonances [Fig.~\ref{fig:Functions}(d)]. Importantly, operation in transmission and reflection mode can be handled in a uniform manner by spectrally interleaving (antimatching) or overlapping (matching) the electric and magnetic resonances, respectively [Fig.~\ref{fig:Functions}(e),(f)].

%\section{Theory of Multiresonant Metasurfaces for a Quadratic Phase Profile}
In order to impart a positive or negative linear chirp (slope of instantaneous frequency) and broaden/compress a broadband Gaussian input pulse, the required response of the metasurface (be it reflection or transmission) should be of the form $H(\omega)=\mathcal{A}\exp\{i[\Phi_2(\omega-\Omega)^2+
\Phi_1(\omega-\Omega)+\Phi_0]\}$, where $\Omega$ is the center frequency of the pulse spectrum and $0<\mathcal{A}\leq 1$ allows for some absorption in a realistic MS. With lowercase $\phi_i \;(i=0,1,2)$ we indicate coefficients of a Taylor expansion about zero frequency instead of $\Omega$; for relations between capital $\Phi_i$ and lowercase $\phi_i$ see Appendix~\ref{sec:Supp:Poles}. Such quadratic transfer functions are used for controlling the group velocity dispersion e.g. in fiber optics \cite{NLFObook}. However, $H(\omega)$ is \emph{not} a physical transfer function (TF) since it does not correspond to a real-valued convolution kernel in the time domain, $h(t)$. In order to obey the required Hermitian symmetry, $|X(\omega)|=|X(-\omega)|$ and $\arg X(\omega)=-\arg X(-\omega)$, we introduce the \emph{symmetrized} transfer function $H'(\omega)=\mathcal{A}\exp\{i\operatorname{sgn}(\omega)[\Phi_2(|\omega|-\Omega)^2+
\Phi_1(|\omega|-\Omega)+\Phi_0)]\}$, denoted by the prime symbol. Using $H'(\omega)$ in the place of $H(\omega)$ introduces a negligible error as long as the signal half-bandwidth is smaller than the central frequency ($\Delta\omega<\Omega$), such that the positive-frequency part $g(\omega)$ of the pulse spectrum $U_\mathrm{in}(\omega)=g(\omega)+g^*(-\omega)$ of the real input signal does not extend into negative frequencies. For the error to be strictly zero, the support of $g(\omega)$ needs to contain only non-negative frequencies, $g(\omega)=0\; \forall\; \omega<0$. For details see Appendix~\ref{sec:Supp:FiniteSupport}.

Although $H'(\omega)$ possesses the correct symmetry, it is discontinuous, i.e., it jumps across the imaginary axis. (In addition, it is not guaranteed to satisfy causality; this will be discussed later on). To side-step this discontinuity, we focus on frequencies $\omega>0$ for which $H'(\omega)$ is meromorphic; this will allow to use the Mittag-Leffler partial fraction expansion of complex analysis \cite{AblowitzFokas}. Note that the analytical continuation of $H'(\omega>0)$ into negative frequencies coincides with the initially defined $H(\omega)$.

\begin{figure}
\centering
\includegraphics[width=8.5cm]{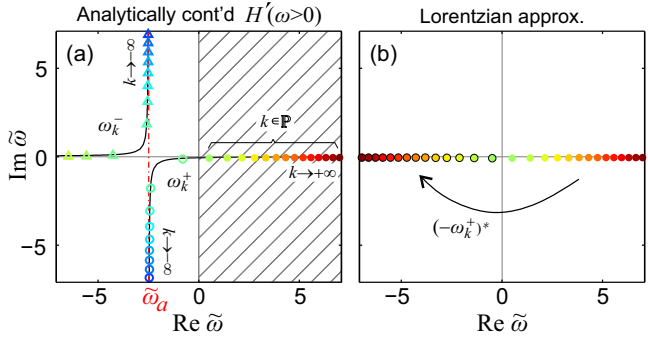}
\caption{\label{fig:ColorCodedPoles} (a)~Positions of poles of Eq.~\eqref{eq:targetSpectrum} in the complex $\omega-$plane assuming the analytically-continued transfer function $H'(\omega>0)$. The two branches, $\omega_k^+$ and $\omega_k^-$, diverge at $\operatorname{Re}(\omega)=\omega_a$. A subset of the poles of the $\omega_k^+$ branch satisfies $\operatorname{Re}\omega_k^+>0$ and $\operatorname{Im}\omega_k^+<0$ and is denoted by $k\in\mathbb{P}$. (b)~Strategy for Lorentzian approximation: the $k\in\mathbb{P}$ poles in panel (a) are used along with their negative-conjugate counterparts. This specific example is for
%$\Phi_2=+0.25$, $\Phi_1=16.62$, $\Phi_0=270$, $\Omega=2\pi\times4.5$ (equivalently,
$\phi_2=+0.25$, $\phi_1=2.48$, $\phi_0=0$, and $\mathcal{A}=0.7$. The normalized frequency is defined as $\tilde{\omega}=\omega\sqrt{|\Phi_2|}$ and the normalized apex frequency equals $\tilde{\omega}_a=-2.48$.
\blue{Note that all the poles in panel (b) satisfy $\operatorname{Im}\omega<0$; the imaginary part is small and they might seem to overlap with the horizontal axis. For the particular poles depicted in panel (b) it holds $-0.0585<\operatorname{Im}\tilde{\omega}<-0.0187$.}}
\end{figure}

We now specify the required surface conductivities of a MS implementing the transfer function $H'(\omega>0)$. For operation in transmission, we require scattering amplitudes $t(\omega)=\mathcal{A}e^{i\phi(\omega)}$ and $r(\omega)=0$, where $\phi(\omega)$ is the quadratic phase. Substituting in the expressions relating plane-wave scattering coefficients with dimensionless conductivities (\blue{$\tilde{\sigma}_{se}=\zeta\sigma_{se}/2$ and $\tilde{\sigma}_{sm}=\sigma_{sm}/(2\zeta)$, where $\zeta$ is the wave impedance}, see Appendix~\ref{sec:Supp:Poles}), we find (for operation in reflection it would  be $\tilde{\sigma}_{sm}=1/\tilde{\sigma}_{se}$)
\begin{equation}\label{eq:targetSpectrum}
\tilde{\sigma}_{se}\!=\tilde{\sigma}_{sm}\!
=\!-i\tan\!\left(\!\frac{\phi(\omega)+i|\log\mathcal{A}|}{2}\!\right)
\!=\!-i\tan z(\omega).
\end{equation}
Equation~\ref{eq:targetSpectrum} constitutes the ``target spectrum'' of the conductivities. However, only certain types of resonant behavior are available in nature. In the following, we will thus be seeking a good approximation of the target spectrum using Lorentzian resonances, which can be physically implemented with resonant meta-atoms. The $\omega-$poles for the desired conductivities of Eq.~\eqref{eq:targetSpectrum} can be found analytically by solving a quadratic equation [see Appendix~\ref{sec:Supp:Poles}], resulting in two sets of poles in the complex $\omega$-plane
\begin{equation}\label{eq:poles}
\begin{split}
\omega_k^{\pm}&=:\omega_a \pm \omega_k
=\Omega-\frac{\Phi_1}{2\Phi_2}\\
&\pm
\sqrt{\left(\frac{\Phi_1}{2\Phi_2}\right)^2
+\frac{1}{\Phi_2}\left[(2k+1)\pi-\Phi_0-i|\log\mathcal{A}|\right]},
\end{split}
\end{equation}
%\begin{equation}\label{eq:poles}
%\begin{split}
%\omega_k^{\pm}&=:\omega_a \pm \omega_k\\
%&=\Omega-\frac{\Phi_1}{2\Phi_2}
%\pm
%\sqrt{\left(\frac{\Phi_1}{2\Phi_2}\right)^2
%+\frac{1}{\Phi_2}\left[(2k+1)\pi-\Phi_0-i|\log\mathcal{A}|\right]},
%\end{split}
%\end{equation}
where $\omega_a$ is a real quantity coinciding with the apex of the parabolic phase and $\omega_k$ a complex quantity that determines the offset of the poles in the complex plane. The poles reside on two curves which asymptotically approach the vertical axis $\operatorname{Re}(\omega)=\omega_a$ for $k\rightarrow-\infty$ and the horizontal axis $\operatorname{Im}(\omega)=0$ for $k\rightarrow+\infty$. Depending on the specific choice for $\Phi_i$, the index $k$ is ``re-normalized'' and the poles shift to different discrete positions along the curves (see Appendix~\ref{sec:Supp:Poles}). The discrete poles along with the underlying continuous curves are depicted in Fig.~\ref{fig:ColorCodedPoles}(a) for a characteristic positive-chirp ($\Phi_2>0$) case.  Note that the pole index $k$ is under the square root, leading to non-uniform spacing along the real axis and a varying imaginary part, in contrast to the case of multiresonant metasurfaces for pulse delay \cite{Tsilipakos:2018}, where the poles are equidistant and the imaginary part constant. The study of pole structure in nanophotonics and metasurfaces/scatterers in particular is recently receiving increased interest, since it can help to achieve advanced functionalities and provide physical insight \cite{Benzaouia:2022,Colom:Arxiv:2022,Wu2020,Gras2019,Grigoriev2013,Valagiannopoulos:2022,Tzortzakakis:2021,Christopoulos:2023}.

In Fig.~\ref{fig:ColorCodedPoles}(a) we have chosen  $\Phi_1>2\Phi_2\Omega$ so that $\omega_a$, where the two branches diverge, lies in the left complex half-plane. When $\omega_a<0$, \emph{all} the poles in the right complex half-plane are predominantly real \emph{and} possess a negative imaginary part. They are compatible with physical resonances and can form the basis for a Lorentzian approximation discussed below [see Fig.~\ref{fig:ColorCodedPoles}(b)]; poles to the left of $\omega_a$ possess a positive imaginary part and would not satisfy causality (anti-resonances). Importantly, this means that there is no fundamental limit on the bandwidth that can be accommodated by the metasurface; in contrast, if $\omega_a>0$ a low-frequency limit for the positive-frequency content of the pulse would be imposed.

Having specified the simple $\omega-$poles of Eq.~\eqref{eq:targetSpectrum}, we can write the corresponding Mittag-Leffler expansion (see Appendix~\ref{sec:Supp:LorApprox})
\begin{equation} \label{eq:PFE}
-i\tan z(\omega)=\sum_{k=-\infty}^{+\infty}
\dfrac{i}{\omega_k\Phi_2}
\left(\dfrac{1}{\omega-\omega_k^{+}}-\dfrac{1}{\omega-\omega_k^{-}}\right).
\end{equation}
The residues are $r_k^+=i/(\omega_k\Phi_2)$ and $r_k^-=-i/(\omega_k\Phi_2)$ for $\omega_k^{+}$ and $\omega_k^{-}$ poles, respectively. We now identify a subset of the poles of the $\omega_k^+$ branch that satisfies $\operatorname{Re}\omega_k^+>0$ and $\operatorname{Im}\omega_k^+<0$ and can play the role of positive-frequency poles of an underdamped linear oscillator (resonant meta-atom), see Appendix~\ref{subsec:Supp:LorApprox1}. The corresponding indices are denoted by $k\in\mathbb{P}$ in Fig.~\ref{fig:ColorCodedPoles}(a). We can thus use these simple poles, along with their complex conjugate counterparts, to construct a physical, Lorentzian approximation of the target spectrum. This procedure is schematically depicted in Fig.~\ref{fig:ColorCodedPoles}(b). Looking at the form of a Lorentzian resonance in the surface conductivity (see Appendix~\ref{subsec:Supp:LorApprox1}), we also require that the corresponding residues are of the form $r_k^+=a(i\omega_k^+)$, with $a\in\mathbb{R}$ and $a>0$. This suggests approximating the actual residues with $r_k^+=i/(\omega_k\Phi_2)
\approx \operatorname{Re}[1/(\omega_k\Phi_2\omega_k^+)]i\omega_k^+$. For any reasonable value of loss, the $k\in \mathbb{P}$  poles are predominantly real and the error of approximating the residues by taking the real part is negligible. The Lorentzian approximation (LA) then takes the form
\begin{equation} \label{eq:LA}
\begin{split}
\tilde{\sigma}_\mathrm{LA}(\omega)\!=\!\!\sum_{k\in\mathbb{P}}
\operatorname{Re}\!\left(\!\dfrac{1}{\omega_k\Phi_2\omega_k^+}\!\!\right)
\!\!\left(\!\dfrac{i\omega_k^{+}}{\omega-\omega_k^{+}}\!
-\!\dfrac{(i\omega_k^{+})^*}{\omega-(-\omega_k^{+})^*}\!\!\right).
\end{split}
\end{equation}
\green{Note that by construction the proposed response function given by Eq.~\eqref{eq:LA} is analytic in the upper half-plane and of Hermitian symmetry (the time-domain counterpart is real). In addition, for a finite number of terms it also holds $\tilde{\sigma}_\mathrm{LA}(\omega)\rightarrow 0$ as $|\omega|\rightarrow\infty$. This means that the real and imaginary parts are related via Kramers-Kronig relations.}
What remains in $\tilde{\sigma}(\omega)=-i\tan z(\omega)=\tilde{\sigma}_\mathrm{LA}(\omega)+\Delta\tilde{\sigma}(\omega)$ is the error of the Lorentzian approximation and is comprised of four contributions: (i)~the subtraction of the negative frequency counterparts we added in Eq.~\eqref{eq:LA}, (ii)~what is left from taking the real part of the residues, (iii)~the poles omitted from the $\omega_k^+$ branch ($k\notin \mathbb{P}$), and (iv)~the entire $\omega_k^-$ branch. See Appendix~\ref{subsec:Supp:LorApprox2} for details.

The procedure is entirely analogous for a negative chirp ($\Phi_2<0$). In this case, necessarily $\omega_a>0$ and only poles in the strip $\operatorname{Re}(\omega)\in(0, \omega_a)$ can be used for the LA; this imposes a high-frequency limit for the positive-frequency content of the pulse [see Appendix, Fig.~\ref{fig:Supp:ColorCodedPoles1}(b)].

\green{
It is also interesting to note that not only the proposed response function, $\tilde{\sigma}_\mathrm{LA}(\omega)$, but also the corresponding transfer function $t=(1-\tilde{\sigma}_\mathrm{LA})/(1+\tilde{\sigma}_\mathrm{LA})$ (we have used $\tilde{\sigma}_{se}(\omega)=\tilde{\sigma}_{sm}(\omega)=\tilde{\sigma}_\mathrm{LA}(\omega)$ in Eq.~\eqref{eq:supp:forwardExprs}, Appendix~\ref{sec:Supp:Poles}) is analytic in the upper half-plane. This is discussed in more detail in Appendix~\ref{sec:Supp:TransfFunctPoles}.
The corresponding scattered field, $t(\omega)-1$ (the total transmitted field is the sum of incident field plus scattered field), possesses the additional property that it vanishes at infinity (for a finite sum of Lorentzians). We thus conclude that Kramers-Kronig relations apply to the scattered field, $t(\omega)-1$.}

%\red{Obviously, since the real and imaginary parts of $t-1$ are related, this will necessarily impose some relation between the corresponding magnitude and phase. In general, it cannot be cast in a simple closed-form relation \cite{Bechhoefer:2011}. However, according to Ref.~\onlinecite{Bechhoefer:2011}, when the function obeys the Kramers-Kronig criteria and moreover does not have zeros in the upper half-plane (as is the case for $t(\omega)-1$, see Fig.~S7(c) of SM), the mag-phase relation can take the form of a Hilbert transform pair (Bode relation).}

\section{Truncation of Infinite Lorentzian Sum and Performance Analysis}

\begin{figure}
\centering
\includegraphics[width=8.5cm]{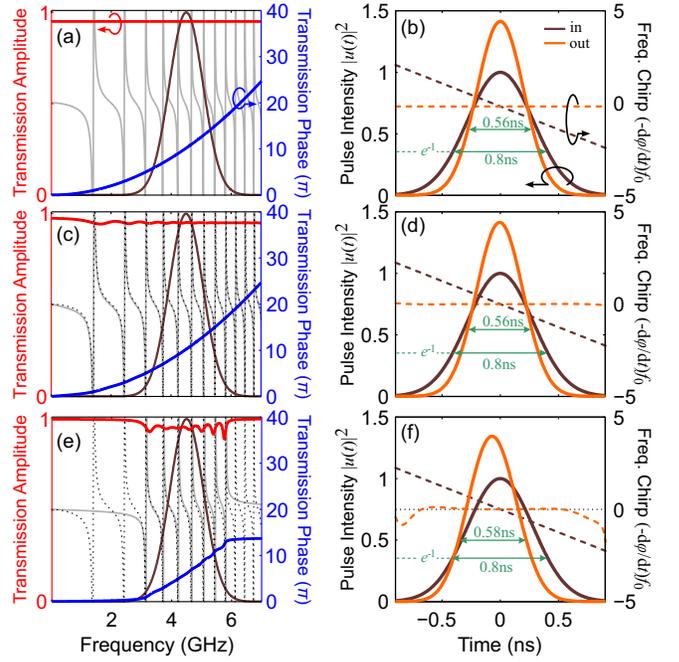}% Here is how to import EPS art
\caption{\label{fig:ResultPosChirp}
Results for operation in transmission. Compression of a negatively pre-chirped ($C<0$) Gaussian broadband pulse upon interaction with a metasurface exhibiting a quadratic phase profile with positive $\Phi_2$ (e.g. for dispersion compensation). Different levels of approximation in the transfer function and impact on the output pulse $u_\mathrm{out}(t)$. (a,b)~Ideal transfer function: (a)~Ideal transmission amplitude and phase; some loss is included ($\mathcal{A}:=|t(\omega)|=0.95$). The corresponding required surface conductivity and the pulse spectrum are overlaid. (b)~Input and output pulse and frequency chirp. The output pulse is de-chirped and compressed by $1/\sqrt{2}$, as verified by the pulse durations measured at the $e^{-1}$ intensity points.  (c,d)~Physical approximation of the ideal target spectrum with an infinite train of Lorentzian resonances [Eq.~\eqref{eq:LA}]: (c)~Transmission amplitude and phase along with surface conductivity and pulse spectrum. The ideal surface conductivity from panel (a) is also included with a dashed line. (d)~Input and output pulse and frequency chirp. The performance is practically indistinguishable from the ideal case. (e,f)~Truncated physical approximation using seven resonances: (e)~Some ripples manifest in the transmission phase/amplitude due to the truncation. The untruncated surface conductivity from panel (c) is included with a dashed line. (f)~The compression is only slightly affected and the residual output chirp is negligible along the duration of the output pulse.}
\end{figure}

\begin{figure}
\centering
\includegraphics[width=8.5cm]{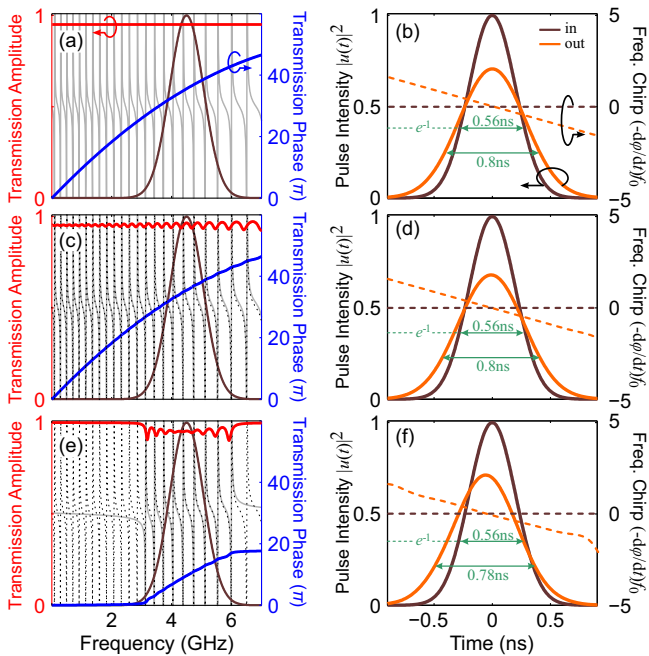}% Here is how to import EPS art
\caption{\label{fig:ResultNegChirp}
Results for operation in transmission. Broadening of an initially un-chirped Gaussian broadband pulse upon interaction with a metasurface exhibiting a quadratic phase profile with negative $\Phi_2$ (e.g. for chirped pulse amplification). Different levels of approximation in the transfer function and impact on the output pulse $u_\mathrm{out}(t)$. (a,b)~Ideal transfer function: (a)~Ideal transmission amplitude and phase; some loss is included ($\mathcal{A}:=|t(\omega)|=0.95$). The corresponding required surface conductivity and the pulse spectrum are overlaid. (b)~Input and output pulse and frequency chirp. The output pulse acquires a negative chirp and is broadened by $\sqrt{2}$, as verified by the pulse durations measured at the $e^{-1}$ intensity points. (c,d)~Physical approximation of the ideal target spectrum with an infinite train of Lorentzian resonances: (c)~Transmission amplitude and phase along with surface conductivity and pulse spectrum. The ideal surface conductivity from panel (a) is also included with a dashed line. (d)~Input and output pulse and frequency chirp. The performance is practically indistinguishable from the ideal case. (e,f)~Truncated physical approximation with nine resonances: (e)~The untruncated surface conductivity from panel (c) is also included with a dashed line. (f)~The broadening is only slightly affected; the output chirp is in good approximation linear along the duration of the output pulse.}
\end{figure}

The final step that enables a practical, physical prescription for the implementation of a metasurface for dispersion compensation and pulse chirping is to truncate the sum in Eq.~\eqref{eq:LA}. The impact of this truncation on the MS performance is tractable and the associated error is negligible provided that the pulse spectrum is accommodated within the bandwidth supplied by the finite set of resonances. This is demonstrated in Fig.~\ref{fig:ResultPosChirp} and \ref{fig:ResultNegChirp}, where the effect of the LA and its truncation on the transfer function of the MS, as well as the pulse in the time domain, are documented.

Figure~\ref{fig:ResultPosChirp} deals with positive chirp ($\Phi_2>0$) and studies compression (dispersion compensation) of a negatively pre-chirped ($C<0$) broadband Gaussian pulse upon interaction with the metasurface. The input pulse is a delayed, modulated Gaussian pulse of the form
$u_\mathrm{in}(t)=\exp[-(1+iC)(t-t_0)^2/(2\tau_0^2)]\exp[-i\Omega(t-t_0)]$, with $\Delta\omega=1/\tau_0$ being the transform limited spectral half-width ($e^{-1}$ intensity point) of the pulse spectrum.
The parameters of the example are:
%\blue{(Grad/s stands for gigaradians per second)}:
\blue{$\Omega=2\pi\times4.5\cdot10^9$~rad/s}, initial chirp $C=-1$, \blue{$\Delta\omega=1/\tau_0=2\pi\times0.28\cdot10^9$~rad/s}, $\mathcal{A}=0.95$, $\Phi_2=+0.04$~ps$^2$, $\Phi_1=2.26$~ps, $\Phi_0=32$ (equivalently, $\phi_2=0.04$~ps$^2$, $\phi_1=0$, $\phi_0=0$).
%The parameters are $\Omega=2\pi\times4.5$~Grad/s, initial chirp $C=-1$, $\Delta\omega=2\pi\times0.28$~Grad/s (transform-limited spectral half-width, $e^{-1}$ intensity point), $\mathcal{A}=0.95$, $\Phi_2=+0.04$~ps$^2$, $\Phi_1=2.26$~ps, $\Phi_0=32$ (equivalently, $\phi_2=0.04$~ps$^2$, $\phi_1=0$, $\phi_0=0$).
Microwave frequencies are selected for this example, since for the physical implementation we can directly rely on an experimentally-verified multiresonant unit cell based on ELC (electric-LC) electric resonators and SRR (split ring resonator) magnetic resonators \cite{Tsilipakos:2021}. The approach of using metallic meta-atoms can be utilized practically unchanged up to THz frequencies. For optical frequencies, Mie resonances in dielectric particles may constitute a favorable approach, since metals are associated with significant resistive loss. Note that such engineering challenges, associated with a particular physical implementation, are outside the scope of the current work, in which we establish the theoretical principles and foundations that are prerequisite to any subsequent physical implementation.
The target spectrum is depicted in Fig.~\ref{fig:ResultPosChirp}(a): the transmission amplitude is flat and equal to $0.95$ over arbitrarily-broad bandwidths and the phase is exactly quadratic. In effect, the input pulse is compressed by exactly $1/\sqrt{2}$, as designed, and the output chirp (variation of instantaneous frequency) is zero across the entire pulse duration [Fig.~\ref{fig:ResultPosChirp}(b)]. The untruncated train of physical Lorentzian resonances [Eq.~\eqref{eq:LA}] is depicted in Fig.~\ref{fig:ResultPosChirp}(c). The target spectrum is included with a dashed line; they are almost indistinguishable and so is the effect on the output pulse [Fig.~\ref{fig:ResultPosChirp}(d)]. Subsequently, the infinite resonance train is truncated keeping only seven resonances [Fig.~\ref{fig:ResultPosChirp}(e)]. The available bandwidth becomes finite but is approximately 3~GHz, corresponding to a vast relative bandwidth of 67\%. Due to the crude truncation, a ripple develops in the transmission amplitude and phase. However, the pulse compression is only slightly affected and the residual output chirp is negligible throughout the duration of the output pulse [Fig.~\ref{fig:ResultPosChirp}(f)]. If even higher integrity is required, one can fine-tune the positions and strengths of the considered resonances after truncation and/or introduce an additional background contribution [see Appendix, Fig.~\ref{fig:Supp:ResFunct1}(d)].

Next, the case of negative chirp ($\Phi_2<0$) and pulse stretching (e.g. for chirped pulse amplification) is considered in Fig.~\ref{fig:ResultNegChirp}. The parameters of the example are: \blue{$\Omega=2\pi\times4.5\cdot10^9$~rad/s}, initial chirp $C=0$, \blue{$\Delta\omega=2\pi\times0.28\cdot10^9$~rad/s} (transform-limited spectral half-width, $e^{-1}$ intensity point), $\mathcal{A}=0.95$, $\Phi_2=-0.04$~ps$^2$, $\Phi_1=2.83$~ps, $\Phi_0=112$ (equivalently, $\phi_2=-0.04$~ps$^2$, $\phi_1=5.1$~ps, $\phi_0=0$). The target spectrum is depicted in Fig.~\ref{fig:ResultNegChirp}(a). The initially un-chirped Gaussian broadband pulse acquires a linear induced chirp and is broadened by a factor $\sqrt{2}$, as designed [Fig.~\ref{fig:ResultNegChirp}(b)]. The infinite Lorentzian approximation is depicted in Fig.~\ref{fig:ResultNegChirp}(c). As was the case with the positive chirp scenario, the output pulse [Fig.~\ref{fig:ResultNegChirp}(d)] is almost indistinguishable compared with the ideal case. Finally, the infinite resonance train is truncated keeping nine resonances [Fig.~\ref{fig:ResultNegChirp}(e)]. Pulse stretching is only slightly affected and output chirp is linear throughout the duration of the output pulse [Fig.~\ref{fig:ResultNegChirp}(f)]. Results for operation in reflection mode (both positive and negative chirp) are included in the Appendix~\ref{sec:Supp:Reflection}.

\section{Conclusion}
In conclusion, we have presented a solution to arbitrarily-strong and arbitrarily-broadband quadratic phase shaping with multiresonant metasurfaces. \blue{Our approach aspires to bring dispersion engineering to the nanoscale and overcome the current limitations of both (i)~conventional, non-resonant approaches with bulk media (too bulky) as well as (ii)~modern, singly-resonant metasurfaces (too narrowband).} The proposed concept is not limited to dispersion compensation \blue{or chirped pulse amplification}, but provides a generic and powerful ultrathin platform for the spatio-temporal control of broadband real-world signals with a myriad of applications in modern optics, microwave photonics, radar and communication systems.

%\vspace{0.3cm}
%\noindent\textbf{\large Acknowledgments}
%\vspace{0.1cm}\\
%\section*{Acknowledgments}
\vspace{0.5cm}
\begin{acknowledgments}
Work at Ames Laboratory was supported by the Department of Energy (Basic Energy Sciences, Division of Materials Sciences and Engineering) under Contract No. DE-AC02-07CH11358.
Support by the Hellenic Foundation for Research and Innovation (H.F.R.I.) under the ``2nd Call for H.F.R.I. Research Projects to support Post-doctoral Researchers'' (Project Number: 916, PHOTOSURF).
\end{acknowledgments}

%\linespread{1.30}
%\renewcommand\baselinestretch{1.3}
\appendix
\onecolumngrid
 
\section{\label{sec:Supp:FiniteSupport}Finding a proper transfer function for pulse chirping}

Let the output physical field quantity be a chirped and delayed modulated Gaussian pulse in the time domain,
\begin{equation}
\begin{split}\label{eq:supp:Etout}
u_\mathrm{out}(t)&=e^{-\displaystyle\frac{(t-t_0)^2}{2\tau^2}}
2\cos\{[\Omega+\Lambda (t-t_0)](t-t_0)+\varphi\}\\
&=\exp\Bigg\{-\left(\frac{1}{2\tau^2}+i\Lambda\right)(t-t_0)^2
-i\Omega(t-t_0)-i\varphi\Bigg\}+c.c.,
\end{split}
\end{equation}
where $\tau>0$ is the output (broadened) pulse duration (i.e., the half-width at the $e^{-1}$ intensity point), $t_0$ is the group delay, $\Omega$ the center frequency, $\Lambda$ a linear chirp of the instantaneous frequency,\footnote{A Gaussian pulse shape maintains its shape and acquires a perfectly linear chirp upon interaction with a dispersive medium possessing a quadratic phase profile \cite{NLFObook}. This is not necessarily true for other pulse shapes.} and $\varphi$ a constant phase that shifts the carrier oscillation with respect to the Gaussian envelope.

In Fourier-space\footnote{$\int_{-\infty}^{+\infty}e^{-(ax^2+bx+c)}dt=
\sqrt{\frac{\pi}{a}}e^{(b^2-4ac)/4a}$. In our case, integrability is guaranteed since $\operatorname{Re}a>0$.}
\begin{equation}
\begin{split}\label{eq:supp:Efout}
U_\mathrm{out}(\omega)&:=\int_{-\infty}^{+\infty}u_\mathrm{out}(t)e^{i\omega t}dt\\&=\sqrt{2\pi}e^{i\omega t_0}\Bigg\{\frac{\tau}{\sqrt{1+2i\Lambda\tau^2}}\;e^{-\displaystyle\frac{1}{2}
\frac{\tau^2}{1+2i\Lambda\tau^2}(\omega-\Omega)^2}e^{-i\varphi} \\
&\quad\quad\quad\quad\quad+\frac{\tau}{\sqrt{1-2i\Lambda\tau^2}}\; e^{-\displaystyle\frac{1}{2}
\frac{\tau^2}{1-2i\Lambda\tau^2}(\omega+\Omega)^2}e^{+i\varphi}\Bigg\}.
\end{split}
\end{equation}
Since a physical field $u(t)$ is real, its Fourier transform should satisfy $U(\omega)=U^*(-\omega)$, i.e., it should be a Hermitian function. This translates into the absolute part being an even, $|U(\omega)|=|U(-\omega)|$, and the argument an odd function of frequency, $\arg U(\omega)=-\arg U(-\omega)$.  In order to define a simple, physical device that would transform between an unchirped and a chirped pulse and vice-versa over a large bandwidth, we aim to separate the output field into an input field times a transfer function (TF) that acts on the phase (which should be of quadratic profile). We can thus write
\begin{equation}
U_\mathrm{out}(\omega)=G(\omega)+G^*(-\omega),
\end{equation}
where
\begin{equation}
\begin{split} \label{eq:supp:Htimesg}
G(\omega)=&H(\omega)\times g(\omega)\\
=&\exp\left(i\{\Lambda(\tau\tau_0)^2(\omega-\Omega)^2+
t_0(\omega-\Omega)-[\varphi-\Omega t_0+\frac{1}{2}\arg(1+i2\Lambda\tau^2)]\}\right)\\
&\times \sqrt{\frac{\tau}{\tau_0}}\sqrt{2\pi\tau_0^2}
\exp\left(-\frac{1}{2}\tau_0^2(\omega-\Omega)^2\right),
\end{split}
\end{equation}
with
\begin{equation} \label{eq:supp:tauVstau0}
\frac{1}{\tau_0^2}:=\frac{1}{\tau^2}+(2\Lambda\tau)^2
\end{equation}
connecting the output ($\tau$) and input ($\tau_0$) pulse durations with the chirp parameter $\Lambda$ (linear slope of instantaneous frequency).\footnote{For a prescribed $\Lambda$ there are two (or none) possible solutions of $\tau$ for a given $\tau_0$. Note, however, that they correspond to a different quadratic coefficient of the spectral phase profile $\Phi_2=\Lambda(\tau\tau_0)^2$.} From Eq.~\eqref{eq:supp:tauVstau0}, one sees that $\tau_0<\tau$ as appropriate due to chirp-induced pulse broadening. The function $g(\omega)$ corresponds to the positive frequency part of the unchirped, undelayed input pulse.
Its spectral half-width at the $e^{-1}$ intensity point is related with the temporal duration through $\Delta\omega=1/\tau_0$, as expected for an unchirped (transform-limited) pulse \cite{NLFObook}.
Note that the inverse Fourier transform $\mathcal{F}^{-1}\{e^{-\frac{1}{2}\tau_0^2\omega^2}\}
=\sqrt{2\pi\tau_0^2}e^{-t^2/(2\tau_0^2)}$; the extra amplitude factor $\sqrt{\tau/\tau_0}>1$ expresses that due to pulse broadening there is a corresponding amplitude decrease in order to preserve the pulse energy; as a result, the input Gaussian pulse possesses a larger amplitude by $\sqrt{\tau/\tau_0}$. Note that this amplitude factor is independent of frequency.

The physical requirement for real fields and a real transfer function means that it should hold
\begin{equation} \label{eq:supp:physreq}
U_\mathrm{out}(\omega)=H(\omega)U_\mathrm{in}(\omega),
\end{equation}
which implies
\begin{equation}
\begin{split} \label{eq:supp:physreq2}
G(\omega)+G^*(-\omega)&=H(\omega)[g(\omega)+g^*(-\omega)]\\
&=H(\omega)g(\omega)+H(\omega)g^*(-\omega)\\
&\overset{!}{=}H(\omega)g(\omega)+H^*(-\omega)g^*(-\omega).
\end{split}
\end{equation}
However, one can readily spot that $H(\omega)\neq H^*(-\omega)$, due to the presence of even powers of $\omega$ in the argument of the transfer function. Hence, $H(\omega)$ in Eq.~\eqref{eq:supp:Htimesg} does \emph{not} represent a physical transfer function.

We now try to find a physical transfer function that transforms the unchirped input pulse, $g(\omega)+g^*(-\omega)$, into the chirped output pulse, $G(\omega)+G^*(-\omega)$. We will demonstrate that this is possible for pulse spectra that do \emph{not} contain zero frequencies. In other words, when the positive-frequency part does not extend into negative frequencies and vice-versa.

Lets assume a modified, \emph{symmetrized} transfer function
\begin{equation} \label{eq:supp:Hmod}
H'(\omega)
=\exp(i\operatorname{sgn}(\omega)\{\Lambda(\tau\tau_0)^2(|\omega|-\Omega)^2+
t_0(|\omega|-\Omega)-[\varphi-\Omega t_0+\frac{1}{2}\arg(1+i2\Lambda\tau^2)]\}),
\end{equation}
where
\begin{equation}
  \operatorname{sgn}(\omega)=
  \begin{cases}+1&:\quad\omega>0\\\phantom{+}0&:\quad\omega=0 \\-1&:\quad\omega<0\end{cases}
\end{equation}
for which the Hermitian symmetry property $H'(\omega)=H'^*(-\omega)$ holds and in addition
\begin{equation}
  H'(\omega)=
  \begin{cases}H(\omega)&:\quad\omega>0\\ H^*(-\omega)&:\quad\omega<0\end{cases}.
\end{equation}
If $\operatorname{supp}\{g(\omega)\}$ excludes frequencies $\omega<0$, then continuing from the last row of Eq.~\eqref{eq:supp:physreq2}
\begin{equation}
  H(\omega)g(\omega)+H^*(-\omega)g^*(-\omega)= H'(\omega)[g(\omega)+g^*(-\omega)]=H'(\omega)g(\omega)+H'^*(-\omega)g^*(-\omega).
\end{equation}
In this case, $H'(\omega)$ is a proper transfer function that maps between input and output pulses.

Next, we want to quantify the error introduced by adopting the modified TF $H'(\omega)$ when the support of $g(\omega)$ is not strictly finite and involves negative frequencies. Let $\Theta(\omega)=\frac{1}{2}(1+\operatorname{sgn}(\omega))$ be the Heaviside function.
%\begin{equation}
%  \operatorname{sgn}(\omega)=
%  \begin{cases}+1,\quad\omega>0\\0,\quad\omega=0 \\-1,\quad\omega<0\end{cases}
%\end{equation}
%be the sign function and
%\begin{equation}
%  \Theta(\omega)=\frac{1}{2}(1+\operatorname{sgn}(\omega))=
%  \begin{cases}+1,\quad\omega>0\\+1/2,\quad\omega=0 \\0,\quad\omega<0\end{cases}
%\end{equation}
%the Heaviside function ($\Theta(\omega)+\Theta(-\omega)\equiv 1$).
Further, let $H_+(\omega):=H(\omega)\Theta(\omega)$ and $H_-(\omega):=H(\omega)\Theta(-\omega)$ be the positive and negative frequency part of $H(\omega)$. Then
\begin{equation} \label{eq:supp:HvsHmod}
  H(\omega)=H_+(\omega)+H_-(\omega)=
  H_+(\omega)+H_+^*(-\omega)-H_+^*(-\omega)+H_-(\omega)=:
  H'(\omega)+P(\omega),
\end{equation}
where it can be seen that
\begin{align}
  H'(\omega)=H(\omega)\Theta(\omega)+H^*(-\omega)\Theta(-\omega)&=
  \begin{cases}H(\omega)&:\quad\omega>0\\ \operatorname{Re}H(\omega)&:\quad\omega=0\\ H^*(-\omega)&:\quad\omega<0\end{cases},\\
  P(\omega)=[H(\omega)-H^*(-\omega)]\Theta(-\omega)&=
  \begin{cases}0&:\quad\omega>0\\ i\operatorname{Im}H(\omega)&:\quad\omega=0\\ H(\omega)-H^*(-\omega)&:\quad\omega<0\end{cases}.
\end{align}

Then, from Eq.~\eqref{eq:supp:HvsHmod} we can write the output pulse spectrum
\begin{equation}
\begin{split}
  H(\omega)g(\omega)+H^*(-\omega)g^*(-\omega) &=H'(\omega)g(\omega)+H'^*(-\omega)g^*(-\omega)\\
  &+P(\omega)g(\omega)+P^*(-\omega)g^*(-\omega),
\end{split}
\end{equation}
which states that if we use the proper transfer function $H'(\omega)$ instead of $H(\omega)$, an error term will be introduced, i.e., the corresponding output spectra differ by the function $E(\omega)$
\begin{equation}
U_\mathrm{out}(\omega) = U'_\mathrm{out}(\omega) + E(\omega).
\end{equation}
Note that the error term is given by
\begin{equation}
  E(\omega)=
  \begin{cases}P^*(-\omega)g^*(-\omega)&:\quad\omega>0\\ -2\operatorname{Im}H(0)\operatorname{Im}g(0)&:\quad\omega=0\\
  P(\omega)g(\omega)&:\quad\omega<0\end{cases}
\end{equation}
and depends \emph{only} on the negative ($\omega<0$) frequency content of $g(\omega)$. (For real input signals $\operatorname{Im}g(0)=0$ resulting in $E(0)=0$). If $g(\omega)=0\; \forall\; \omega<0$, a quite typical case for signals where the pulse half-bandwidth is smaller than the center frequency ($\Delta\omega<\Omega$), then the error vanishes.

\begin{figure*}[h!]
\includegraphics{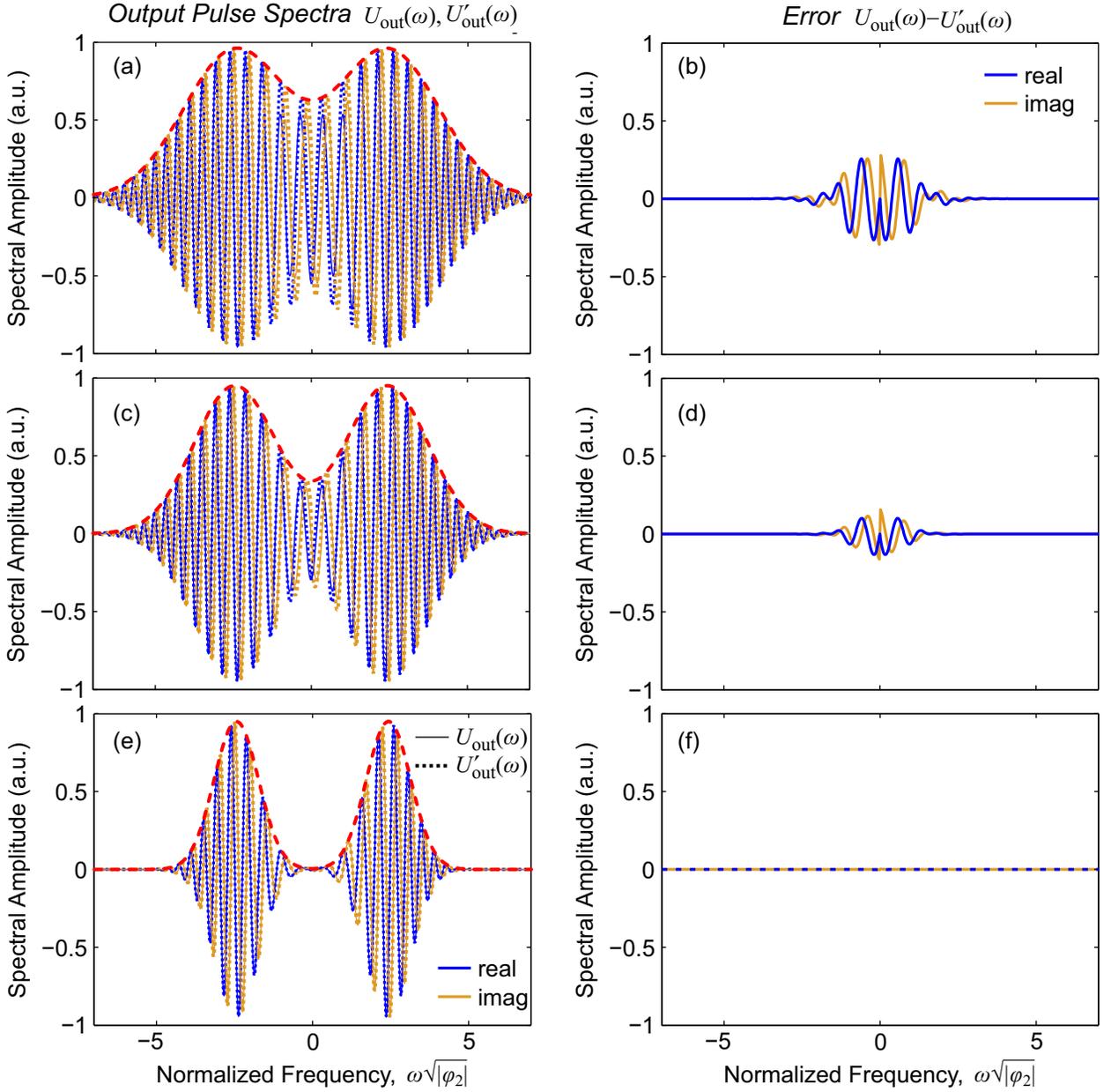}
\caption{\label{fig:Supp:ExcludeZeroFreq} Comparison of output pulse spectra when using the initially considered vs the symmetrized transfer functions, $U_\mathrm{out}(\omega)=H(\omega)U_\mathrm{in}(\omega)$ (solid curves) vs $U'_\mathrm{out}(\omega)=H'(\omega)U_\mathrm{in}(\omega)$ (dashed curves), for input pulses whose positive frequency part extends to negative frequencies ($\Delta\omega\sim\Omega$). The envelope of the input pulse spectrum is included with a red dashed line. The error (difference $U_\mathrm{out}(\omega)-U'_\mathrm{out}(\omega)$) is plotted on the right panels. Progressively, the input pulse bandwidth is decreased, the positive frequency part crosses less into negative frequencies and the error becomes negligible. The parameters of the example are: $\Phi_2=+0.15$, $\Phi_1=5$, $\Phi_0=30$, $\mathcal{A}=0.95$, pulse center frequency $\Omega=2\pi$ ($\tilde{\Omega}=2.43)$, and input pulse spectral half-width ($e^{-1}$ intensity point) $\Delta\omega=4.24, 3.39, 1.83$ for the three cases ($\Delta\tilde{\omega}=1.64, 1.31, 0.71$).}
\end{figure*}

In Fig.~\ref{fig:Supp:ExcludeZeroFreq} we plot the output pulse spectra, $U_\mathrm{out}(\omega)$ and $U'_\mathrm{out}(\omega)$, when using the initially considered, $H(\omega)$, or symmetrized, $H'(\omega)$, transfer functions, respectively. Their difference, i.e., the error term, is plotted on the right panels. As the pulse bandwidth decreases, the positive frequency part crosses less into negative frequencies and the error becomes negligible.

Note that the symmetrized transfer function is physical in the sense that it corresponds to a real-valued convolution kernel. However, it is discontinuous, i.e., it jumps across the imaginary axis, as seen by the behavior of the error in the right panels of Fig.~\ref{fig:Supp:ExcludeZeroFreq}. In order to satisfy causality, an additional requirement should hold: the poles of $H'(\omega)$ should lie in the lower complex half-plane. This will be discussed in the following Sections of the Appendix. The initially considered transfer function can be thought as the analytical continuation of $H'(\omega>0)$ into the negative frequencies; this  removes the discontinuity but the resulting TF does not possess the correct symmetry.

\section{\label{sec:Supp:Poles}Analytic structure of sheet conductivities that implement the transfer function}
% Analytic solution for poles and visualization in the complex plane

Let us consider an abstract metasurface, with which we aim to implement the pulse-chirping transfer function derived in Section~\ref{sec:Supp:FiniteSupport}. The metasurface is described by electric and magnetic complex surface conductivities ${\sigma}_{se}$ and ${\sigma}_{sm}$ measured in $\mathrm{S}$ and $\Omega$, respectively.\footnote{These quantities are related to the surface susceptibilities through ${\sigma}_{se}=-i\omega\varepsilon_0\chi_{se}$, ${\sigma}_{sm}=-i\omega\mu_0\chi_{sm}$ and can be also found as $Y_{se}$ and $Z_{sm}$ in the literature.} %\cite{Pfeiffer:2013}
The equations relating the surface conductivities with reflection and transmission plane wave scattering coefficients are \cite{Tsilipakos:2018} %,Tassin:2012,Holloway:2011
\begin{subequations}\label{eq:supp:forwardExprs}
\begin{align}
\label{eq:supp:r}r(\omega,\theta)&=\frac{-\tilde{\sigma}_{se}+\tilde{\sigma}_{sm}}{1+ \tilde{\sigma}_{se}\tilde{\sigma}_{sm}+\tilde{\sigma}_{se}+\tilde{\sigma}_{sm}},\\
\label{eq:supp:t}t(\omega,\theta)&=\frac{1-\tilde{\sigma}_{se}\tilde{\sigma}_{sm}}{1+ \tilde{\sigma}_{se}\tilde{\sigma}_{sm}+\tilde{\sigma}_{se}+\tilde{\sigma}_{sm}},
\end{align}
\end{subequations}
\begin{subequations} \label{eq:supp:inverseExprs}
\begin{align}
\tilde{\sigma}_{se}(\omega,\theta)=\frac{\zeta\sigma_{se}}{2}&= \frac{1-r-t}{1+r+t},\\
\tilde{\sigma}_{sm}(\omega,\theta)=\frac{\sigma_{sm}}{2\zeta}&= \frac{1+r-t}{1-r+t},
\end{align}
\end{subequations}
where we have defined dimensionless conductivities $\tilde{\sigma}_{se}(\omega,\theta)=\zeta\sigma_{se}/2$ and $\tilde{\sigma}_{sm}(\omega,\theta)=\sigma_{sm}/(2\zeta)$. Note that
$\zeta^\mathrm{TE}(\theta)=\omega\mu/k_\perp=\eta\sec(\theta)$ and $\zeta^\mathrm{TM}(\theta)=k_\perp/(\omega\varepsilon)=\eta\cos(\theta)$ for the TE and TM polarization, respectively, where $\theta$ is the incidence angle and $\eta=\sqrt{\mu/\varepsilon}$ is the characteristic impedance of the homogeneous host medium.

For operation in transmission, the prescription for the scattering amplitudes should be $t(\omega)=\mathcal{A}e^{i\phi(\omega)}$ and $r(\omega)=0$, where $\phi(\omega)$ is the quadratic phase of the transfer function defined in Eq.~\eqref{eq:supp:Hmod} and $0<\mathcal{A}\leq 1$ allows for absorption in the metasurface. Substituting in Eq.~\eqref{eq:supp:inverseExprs}, the required metasurface conductivity (i.e., the target spectrum) is\footnote{For operation in reflection it would simply be $\tilde{\sigma}_{sm}=1/\tilde{\sigma}_{se}
=(1-\mathcal{A}e^{i\phi(\omega)})/(1+\mathcal{A}e^{i\phi(\omega)})$.}
\begin{equation}\label{eq:supp:targetSpectrum}
\begin{split}
\tilde{\sigma}_{se}=\tilde{\sigma}_{sm}&= \frac{1-\mathcal{A}e^{i\phi(\omega)}}{1+\mathcal{A}e^{i\phi(\omega)}}
=\frac{1-e^{i[\phi(\omega)-i\log\mathcal{A}]}}{1+e^{i[\phi(\omega)-i\log\mathcal{A}]}}\\
&=-i\tan(\frac{\phi(\omega)+i|\log\mathcal{A}|}{2})\\
&=-i\tan z(\omega).
\end{split}
\end{equation}
The phase profile in Eq.~\eqref{eq:supp:targetSpectrum}, $\phi(\omega)$, should be quadratic to allow for pulse chirping and dispersion compensation. According to the discussion in Section~\ref{sec:Supp:FiniteSupport}, we can write the complex function $z(\omega)$ in the following form that will result in the correct symmetry for the transfer function $t(\omega)$
\begin{equation}\label{eq:supp:z}
z(\omega)= \frac{1}{2}\operatorname{sgn}(\omega) \left[\Phi_2(|\omega|-\Omega)^2+
\Phi_1(|\omega|-\Omega)+\Phi_0)\right]+i\frac{1}{2}|\log\mathcal{A}|.
\end{equation}
%We will use $\epsilon=|\log\mathcal{A}|/2$ henceforth to simplify the expression and denote that in general loss is desired to be small; however, we do not make any assumption on the level of loss in what follows.
Comparing with Eq.~\eqref{eq:supp:Htimesg}, we identify
\begin{equation}
\begin{split}
\Phi_2&=\Lambda(\tau\tau_0)^2,\\
\Phi_1&=t_0,\\
\Phi_0&=-[\varphi-\Omega t_0+\frac{1}{2}\arg(1+i2\Lambda\tau^2)].
\end{split}
\end{equation}
It is also useful to note that if we use lowercase $\phi_i \;(i=0,1,2)$ to indicate a Taylor expansion of the phase about zero frequency instead of $\Omega$, i.e., $\phi(\omega)=\operatorname{sgn}(\omega)[\phi_2|\omega|^2+
\phi_1|\omega|+\phi_0]$, the following relations hold
\begin{equation} \label{eq:supp:PhiVsphi}
\begin{aligned}
&\phi_2=\Phi_2\\
&\phi_1=\Phi_1-2\Phi_2\Omega\\
&\phi_0=\Phi_0-\Phi_1\Omega+\Phi_2\Omega^2
\end{aligned}
\quad\;\Leftrightarrow\quad\;
\begin{aligned}
&\Phi_2=\phi_2\\
&\Phi_1=\phi_1+2\phi_2\Omega\\
&\Phi_0=\phi_0+\phi_1\Omega+\phi_2\Omega^2
\end{aligned}.
\end{equation}

Equation~\eqref{eq:supp:targetSpectrum} constitutes the target spectrum of the conductivities. However, only certain types of resonant behavior are available in nature. In the following, we will thus be seeking a good approximation of the target spectrum using Lorentzian resonances, which can be provided by subwavelength meta-atoms. Now let $\Phi_2\neq0$ and focus on $\omega>0$, i.e., on the analytical continuation of $H'(\omega)$ for $\omega>0$ into negative frequencies (this coincides with $H(\omega)$, see Section~\ref{sec:Supp:FiniteSupport}). The poles of $H'(\omega>0)$ can be found by solving the quadratic equation
\begin{equation}\label{eq:supp:quadrEq}
z(\omega)=\frac{1}{2}\left[\Phi_2(\omega-\Omega)^2+
\Phi_1(\omega-\Omega)+\Phi_0+i|\log\mathcal{A}|)\right]=k\pi+\frac{\pi}{2}, \quad k\in \mathbb{Z},
\end{equation}
hence,
\begin{equation}\label{eq:supp:quadrEq2}
(\omega-\Omega)^2 + \frac{\Phi_1}{\Phi_2}(\omega-\Omega) - \frac{1}{\Phi_2}[(2k+1)\pi-\Phi_0-i|\log\mathcal{A}|]=0, \quad k\in \mathbb{Z},
\end{equation}
resulting in two sets of poles in the complex $\omega$-plane, $\omega_k^{\pm}$, marked by the sign selection and given by
\begin{equation}\label{eq:supp:poles}
\begin{split}
\omega_k^{\pm}&=\Omega-\frac{\Phi_1}{2\Phi_2}\pm
\sqrt{\left(\frac{\Phi_1}{2\Phi_2}\right)^2
+\frac{1}{\Phi_2}\left[(2k+1)\pi-\Phi_0-i|\log\mathcal{A}|\right]}\\
%&=\Omega-\frac{\Phi_1}{2\Phi_2}\pm
%\sqrt{\left(\frac{\Phi_1}{2\Phi_2}\right)^2+\frac{1}{\Phi_2}q_k}\\
&=:\omega_a \pm \omega_k,
\end{split}
\end{equation}
where $\omega_a=\Omega-\Phi_1/(2\Phi_2)=-\phi_1/(2\phi_2)$ is a real quantity denoting the apex of the parabolic phase and $\omega_k$ is a complex quantity that determines the offset of the poles in the complex plane about the vertical axis $\operatorname{Re}(\omega)=\omega_a$; it depends on the pole index $k$. The quantity $q_k=(2k+1)\pi-\Phi_0-i|\log\mathcal{A}|$ inside the square bracket in Eq.~\eqref{eq:supp:poles} is connected with the poles of a metasurface for the simpler case of broadband group delay (linear phase profile) \cite{Tsilipakos:2018,Tsilipakos:2020}. In the group delay case, the poles are equidistantly spaced along the real axis and their imaginary part is constant. In the present case (quadratic phase profile), this quantity is under a square root: this leads to uneven spacing of the poles along the real axis and a varying imaginary part.
%It also helps to normalize with $\sqrt{|\phi_2|}$ resulting in
%\begin{equation}\label{eq:supp:polesNorm}
%\begin{split}
%\tilde{\omega}_k^{\pm}=\omega_k^{\pm}\sqrt{|\phi_2|}
%&=\tilde{\Omega}-\operatorname{sgn}\{\Phi_2\}\frac{\Phi_1}{2\sqrt{|\Phi_2|}}\pm
%\sqrt{\left(\frac{\Phi_1}{2}\right)^2-\operatorname{sgn}\{\Phi_2\}q_k}\\
%&=\tilde{\omega}_a \pm \tilde{\omega}_k,
%\end{split}
%\end{equation}
In terms of the lowercase $\phi_i$ we can write [Eq.~\eqref{eq:supp:PhiVsphi}]
\begin{equation}\label{eq:supp:poles2}
\begin{split}
\omega_k^{\pm}&=-\frac{\phi_1}{2\phi_2}\pm
\sqrt{\left(\frac{\phi_1}{2\phi_2}\right)^2+\frac{1}{\phi_2}
[(2k+1)\pi-\phi_0-i|\log\mathcal{A}|]}\\
&=\omega_a \pm \omega_k,
\end{split}
\end{equation}

From Eq.~\eqref{eq:supp:poles} we expect two branches of discrete poles that shift between being predominantly real or predominantly imaginary depending on the index $k$ and whether the quantity underneath the square root is positive or negative. We can cast $\omega_k$ in the form
\begin{equation}\label{eq:supp:wk}
\omega_k = \sqrt{\frac{2\pi}{\Phi_2}\left\{k+\frac{\Phi_2}{2\pi}\left[ \left(\frac{\Phi_1}{2\Phi_2}\right)^2 - \frac{\Phi_0}{\Phi_2} \right]+\frac{1}{2}\right\}
-\frac{i|\log\mathcal{A}|}{\Phi_2}},\quad k\in\mathbb{Z}\\
\end{equation}
and see that the quantity in the square bracket which depends on the selected $\Phi_2$, $\Phi_1$, $\Phi_0$ ``re-normalizes'' the index $k$ shifting the discrete subset $k\in\mathbb{Z}$ within the continuum $k\in\mathbb{R}$. Hence, for any choice of $\Phi_2$, $\Phi_1$, $\Phi_0$ the poles
$\omega_k^{\pm}$ are discrete points on the curves
\begin{equation}\label{eq:supp:curves}
\omega^{\pm}(\kappa) = \omega_a \pm \sqrt{\frac{2\pi}{\Phi_2}\left(\kappa+\frac{1}{2}\right)
-\frac{i|\log\mathcal{A}|}{\Phi_2}},\quad \kappa\in\mathbb{R}
\end{equation}

\begin{figure*}[h]
\includegraphics[width=14cm]{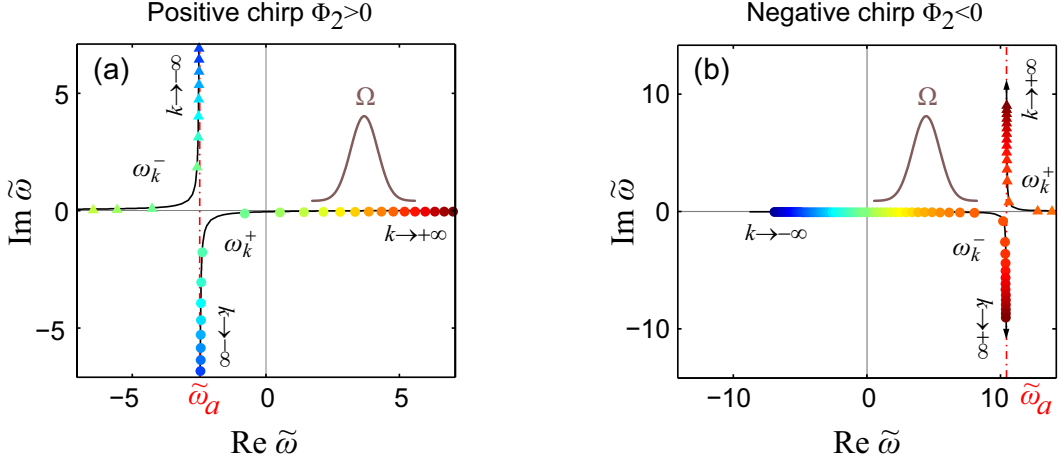}
\caption{\label{fig:Supp:ColorCodedPoles1} Pole positions for the conductivity in Eq.~\eqref{eq:supp:targetSpectrum} assuming the analytically-continued transfer function $H'(\omega>0)$ for (a) a positive ($\Phi_2>0$) and (b) a negative ($\Phi_2<0$) chirp scenario. The frequency is normalized $\tilde{\omega}=\omega\sqrt{|\Phi_2|}$. The two branches are clearly marked. Color-coding corresponds to the value of the $k$ index (red: $k\rightarrow+\infty$, blue: $k\rightarrow-\infty$). Parameters for (a): %$\Phi_2=+0.25$, $\Phi_1=16.62$, $\Phi_0=270$, $\Omega=2\pi\times4.5$ (equivalently,
$\phi_2=+0.25$, $\phi_1=2.48$, $\phi_0=0$, and $\mathcal{A}=0.7$. the normalized apex frequency equals $\tilde{\omega}_a=-2.48$. Parameters for (b):
%$\Phi_2=+0.25$, $\Phi_1=3.68$, $\Phi_0=96$, $\Omega=2\pi\times4.5$ (equivalently,
$\phi_2=-0.25$, $\phi_1=10.45$, $\phi_0=0$, and $\mathcal{A}=0.7$. The normalized apex frequency equals $\tilde{\omega}_a=+10.45$.}
\end{figure*}

The two branches of discrete poles along with the underlying continuous curves are depicted in Fig.~\ref{fig:Supp:ColorCodedPoles1} for both a positive ($\Phi_2>0$) and a negative ($\Phi_2<0$) chirp scenario. For the positive chirp scenario [Fig.~\ref{fig:Supp:ColorCodedPoles1}(a)], $\Phi_1/(2\Phi_2)>0$ and the apex frequency $\omega_a=\Omega-\Phi_1/(2\Phi_2)$ where the two branches diverge lies in the left complex half-plane when $\Phi_1>2\Phi_2\Omega$. This is desirable in order to stay away from the pulse bandwidth (positive frequency part). Note that poles to the left of $\omega_a$ possess a positive imaginary part and would not satisfy causality. For the negative chirp scenario [Fig.~\ref{fig:Supp:ColorCodedPoles1}(b)], the apex frequency $\omega_a=\Omega-\Phi_1/(2\Phi_2)$ lies always in the right complex half-plane since $\Phi_2<0$ and $\Phi_1$ should be positive. It is, thus, required to position $\omega_a$ at a sufficiently high frequency to avoid running into poles that lie to the right of $\omega_a$ and possess a positive imaginary part.

Let us now visualize the positions of the poles by evaluating the conductivity in the complex frequency plane and plotting the magnitude (absolute part). We will compare (i)~the symmetrized TF $H'(\omega)$ that possesses the correct symmetry (Hermitian), but jumps at the imaginary axis (ii)~the analytically continued $H'(\omega>0)\equiv H(\omega)$ that is meromorphic in the entire $\omega$-plane but does not possess the correct symmetry.

\begin{figure*}[h!]
\includegraphics{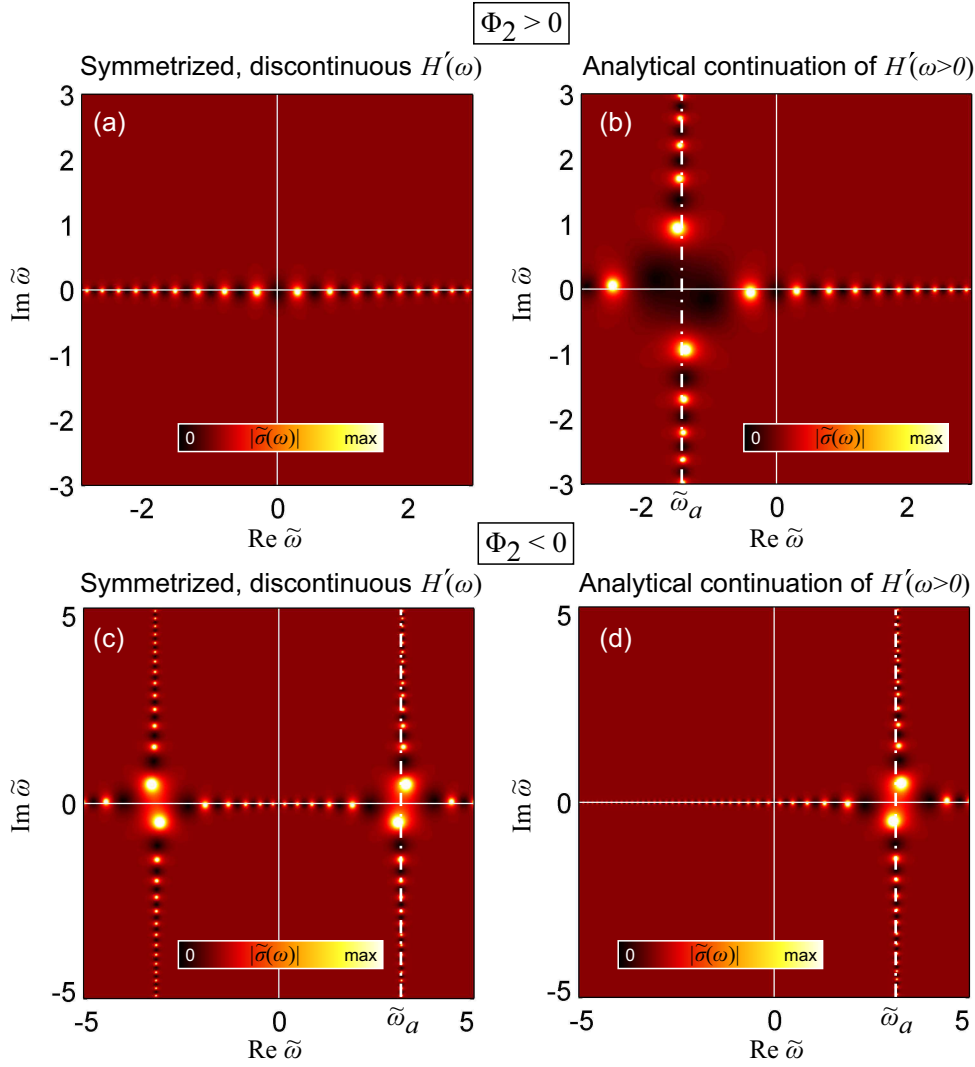}
\caption{\label{fig:Supp:PcolorPoles} Comparing the pole positions of the conductivity corresponding to the transfer function $H'(\omega)$ and the analytically continued $H'(\omega>0) \equiv H(\omega)$. The magnitude (absolute value) of the complex conductivity is plotted in the complex frequency plane $\tilde{\omega}=\omega\sqrt{|\Phi_2|}$. (a,b)~Positive chirp case with $\phi_1=5.16\sqrt{\phi_2}$,  $\phi_0=0$, and $\mathcal{A}=0.7$. (c,d)~Negative chirp case with $\phi_1=11.62\sqrt{|\phi_2|}$,  $\phi_0=0$, and $\mathcal{A}=0.7$.}
\end{figure*}

The positive chirp scenario is depicted in Fig.~\ref{fig:Supp:PcolorPoles}(a),(b). When $\omega_a<0$, \emph{all} the poles in the right complex half-plane are predominantly-real \emph{and} possess a negative imaginary part. These poles can form the basis for a Lorentzian approximation discussed in Section~\ref{sec:Supp:LorApprox}. %therefore, the corresponding $k$ indices will be denoted as $k\in\mathbb{P}$, a subset of $k\in\mathbb{Z}$ for the $\omega_k^+$ branch.
Note that the symmetrized $H'(\omega)$ does not lead to the diverging behavior of the pole positions, since the negative frequencies are removed and replaced by a different analytic function; however, if $\omega_a>0$ ($\Phi_1<2\Phi_2\Omega$) two such divergences would occur: at $\omega_a$ and $-\omega_a$. In terms of the Lorentzian approximation that will follow (Section~\ref{sec:Supp:LorApprox}), this would impose a low-frequency limit for the positive-frequency content of the pulse, $g(\omega)$.

The negative chirp scenario is depicted in Fig.~\ref{fig:Supp:PcolorPoles}(c),(d). In this case, necessarily $\omega_a>0$ and two pole divergences are seen when using the symmetrized $H'(\omega)$. Only poles inside $(-\omega_a, \omega_a)$ are predominantly real \emph{and} possess a negative imaginary part.  In terms of the Lorentzian approximation that will follow (Section~\ref{sec:Supp:LorApprox}), this imposes a high-frequency limit for the positive-frequency content of the pulse, $g(\omega)$.
%A subset of the poles in the  $\omega_k^-$ branch (i) are predominantly real and feature a (ii) positive real part and (iii) negative imaginary part. They will be used for the Lorentzian approximation and the corresponding indices are denoted with $k\in\mathbb{P}$.

\section{\label{sec:Supp:LorApprox}Partial fraction expansion and multi-resonant Lorentzian approximation}

We next perform a partial fraction expansion of Eq.~\eqref{eq:supp:targetSpectrum}, in order to proceed to approximate the target spectrum for the surface conductivities by a physically-realizable train of Lorentzian resonances. The Mittag-Leffler expansion\footnote{Expansion for meromorphic functions into (many) simple poles \cite{AblowitzFokas}.}  of $f(z(\omega))=-i\tan z(\omega)$ reads
\begin{equation} \label{eq:supp:ML}
\begin{split}
f(z(\omega))&= f(0)+\sum_{k=-\infty}^{+\infty}\dfrac{r_k}{z_k}
+\sum_{k=-\infty}^{+\infty}\dfrac{r_k}{z-z_k}\\
&=i\sum_{k=-\infty}^{+\infty}\dfrac{1}{z-z_k},\quad k\in\mathbb{Z},
\end{split}
\end{equation}
where (i)~$z_k=k\pi+\pi/2$ are the poles of $\tan(z)$, (ii)~it is easy to identify that the residues at the poles of $f(z)$ equal $r_k=i$, and (iii)~the first and second terms in the first row of Eq.~\eqref{eq:supp:ML} amount to zero. It now helps to write the denominator (second order equation of $\omega$) in a factored form in terms of the $\omega$-poles identified in Section~\ref{sec:Supp:Poles}:
\begin{equation}
\begin{split}
z(\omega)-z_k &=
\frac{1}{2} \left[\Phi_2(\omega-\Omega)^2+
\Phi_1(\omega-\Omega)+\Phi_0)\right]+i\frac{|\log\mathcal{A}|}{2} - z_k \\
&=\frac{\Phi_2}{2}(\omega-\omega_k^{+})(\omega-\omega_k^{-}).
\end{split}
\end{equation}
Then, we can write
\begin{equation} \label{eq:supp:PFE}
\begin{split}
-i\tan z(\omega)&=i\frac{2}{\Phi_2}\sum_{k=-\infty}^{+\infty}
\dfrac{1}{(\omega-\omega_k^{+})(\omega-\omega_k^{-})}\\
&=i\frac{2}{\Phi_2}\sum_{k=-\infty}^{+\infty}
\dfrac{1}{\omega_k^{+}-\omega_k^{-}}
\left(\dfrac{1}{\omega-\omega_k^{+}}-\dfrac{1}{\omega-\omega_k^{-}}\right)\\
&=\sum_{k=-\infty}^{+\infty}
\dfrac{i}{\omega_k\Phi_2}
\left(\dfrac{1}{\omega-\omega_k^{+}}-\dfrac{1}{\omega-\omega_k^{-}}\right),
\end{split}
\end{equation}
which states that we have expanded $\tilde{\sigma}(\omega)=-i\tan z(\omega)$ in partial fractions of simple poles, $\omega_k^{+}$ and $\omega_k^{-}$, and allows to directly identify the residues, i.e., $r_k^+=i/(\omega_k\Phi_2)$ and $r_k^-=-i/(\omega_k\Phi_2)$.

\subsection{Dispersion of a Lorentzian meta-atom resonance\label{subsec:Supp:LorApprox1}}
Let us consider a homogenizable metamaterial comprised of resonant meta-atoms. The susceptibility of such a linear driven oscillator exhibits a dispersion of the following form
\begin{equation}
\begin{split}
\chi_L(\omega)&=-\dfrac{\omega_0^2 \chi_0}{\omega^2-\omega_0^2+i\Gamma\omega}\quad;\quad \omega_0, \chi_0, \Gamma>0\\
&=-\dfrac{\omega_0^2 \chi_0}{(\omega-\omega^+)(\omega-\omega^-)},
\end{split}
\end{equation}
where
\begin{equation} \label{eq:supp:OscPoles}
\omega^\pm=-\dfrac{i\Gamma}{2}\pm
\sqrt{\omega_0^2-\left(\dfrac{\Gamma}{2}\right)^2}.
\end{equation}
Note that for an underdamped oscillator ($\omega_0^2-(\Gamma/2)^2>0$) it holds $\operatorname{Re}\omega^+>0$, $\operatorname{Im}\omega^+<0$, and $\omega^\pm=(-\omega^\mp)^*$. The latter can be used to show that $\chi_L(-\omega)=\chi_L^*(\omega)$, as anticipated for a physical, real-valued polarization in the time domain.

The corresponding surface conductivity of a metasurface (a thin sheet of thickness $d$ of this metamaterial) is
\begin{equation}
\begin{split}
\sigma_L(\omega)&=-i\omega\varepsilon_0\chi_L(\omega)d\\
&=+\dfrac{i\varepsilon_0\omega_0^2\chi_0d}{\omega^+ - \omega^-}
\left(\dfrac{\omega^+}{\omega-\omega^+} - \dfrac{\omega^-}{\omega-\omega^-}\right).
\end{split}
\end{equation}
It can be seen that the conductivity for a physically-realizable Lorentzian has two poles, $\omega^+$ and $\omega^-=(-\omega^+)^*$, and the corresponding residues are
\begin{equation} \label{eq:supp:OscResidues}
\begin{split}
r^+&=i\varepsilon_0\omega_0^2\chi_0d\dfrac{\omega^+}{\omega^++(\omega^+)^*}
=\varepsilon_0\omega_0^2\chi_0d\dfrac{i\omega^+}{2\operatorname{Re}\omega^+},\\
r^-&=i\varepsilon_0\omega_0^2\chi_0d\dfrac{-(-\omega^+)^*}{\omega^++(\omega^+)^*}
=-(r^+)^*.
\end{split}
\end{equation}
This poses a constraint on the poles and residues of any resonant term in the multi-resonant expansion of the metasurface sheet conductivities that can be implemented by a real physical linear resonant meta-atom.

%therefore, the corresponding $k$ indices will be denoted as $k\in\mathbb{P}$, a subset of $k\in\mathbb{Z}$ for the $\omega_k^+$ branch.

\subsection{Lorentzian approximation of target spectrum for surface conductivity\label{subsec:Supp:LorApprox2}}
\begin{figure*}[h]
\includegraphics{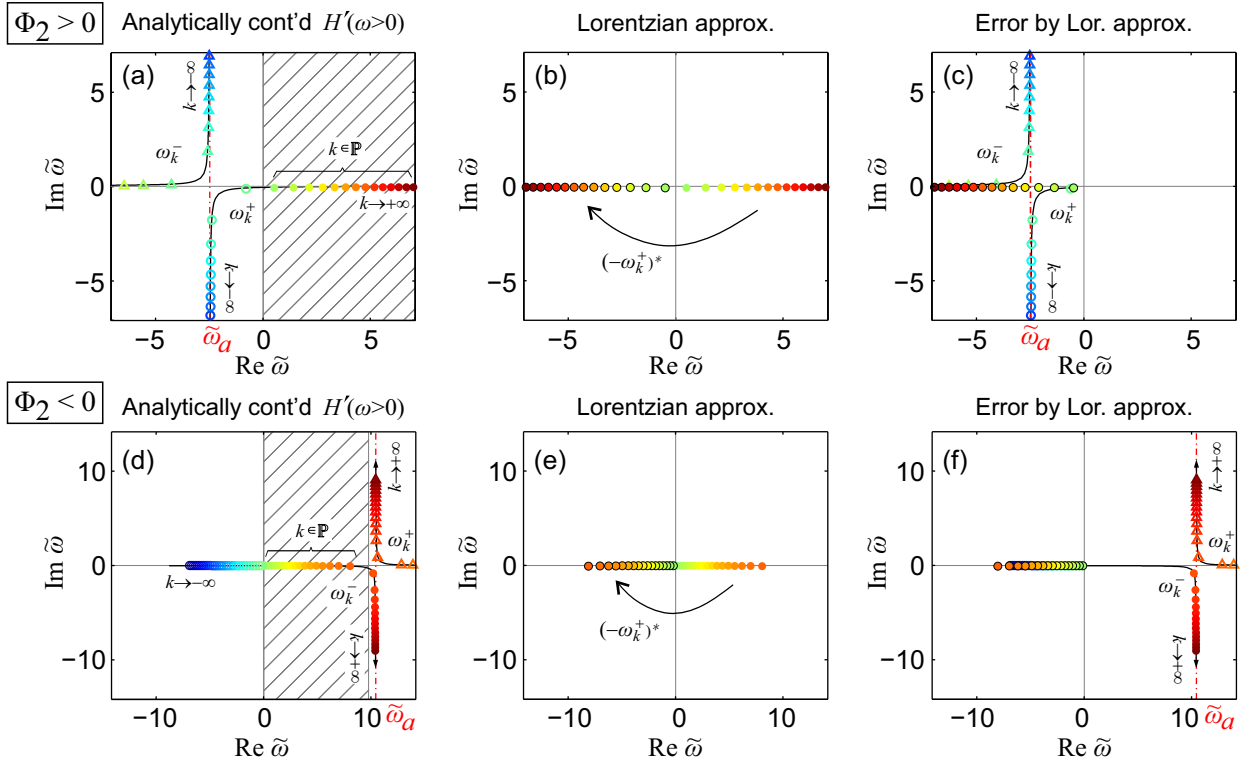}
\caption{\label{fig:Supp:ColorCodedPoles2} (a,b,c) \emph{Positive} chirp case with parameters $\phi_2=+0.25$, $\phi_1=2.48$, $\phi_0=0$, and $\mathcal{A}=0.7$. The frequency is normalized according to $\tilde{\omega}=\omega\sqrt{|\Phi_2|}$ and the normalized apex frequency equals $\tilde{\omega}_a=-2.48$. Positions of conductivity poles for (a) the analytically-continued symmetrized transfer function $H'(\omega>0)$ the (b) Lorentzian approximation (the positive half-plane poles in panel (a) are used along with their negative-conjugate counterparts) and (c) the error term by the Lorentzian approximation. (d,e,f) \emph{Negative} chirp case with
$\phi_2=-0.25$, $\phi_1=10.45$, $\phi_0=0$, and $\mathcal{A}=0.7$. The normalized apex frequency equals $\tilde{\omega}_a=+10.45$. Positions of conductivity poles for (d) the analytically-continued symmetrized transfer function $H'(\omega>0)$ the (e) Lorentzian approximation (the positive half-plane poles in panel (d) are used along with their negative-conjugate counterparts) and (f) the error term by the Lorentzian approximation.}
\end{figure*}

We can now construct a Lorentzian approximation to the target spectrum of Eq.~\eqref{eq:supp:PFE}. Let us first focus on the positive chirp case, $\Phi_2>0$. Observing Fig.~\ref{fig:Supp:ColorCodedPoles1}(a), only a subset of the poles of the $\omega_k^+$ branch  satisfies $\operatorname{Re}\omega_k^+>0$ and $\operatorname{Im}\omega_k^+<0$ and can play the role of positive-frequency poles of a dampened oscillator [cf. Eq.~\eqref{eq:supp:OscPoles}]. The corresponding indices will be denoted by $k\in\mathbb{P}$ [see Fig.~\ref{fig:Supp:ColorCodedPoles2}(a)]. Looking additionally at the oscillator residues [Eq.~\eqref{eq:supp:OscResidues}] we require $r_k^+$ to be of the form $r_k^+=a(i\omega_k^+)$, with $a\in\mathbb{R}$ and $a>0$ [$a$ corresponds to $\varepsilon_0\omega_0^2\chi_0d/(2\operatorname{Re}\omega^+)$ in Eq.~\eqref{eq:supp:OscResidues}]. This suggests to approximate the actual residues with
\begin{equation} \label{eq:supp:Residues}
\begin{split}
r_k^+&=\dfrac{i}{\omega_k\Phi_2}=\dfrac{i\omega_k^+}{\omega_k\Phi_2\omega_k^+}
\approx \operatorname{Re}\left(\dfrac{1}{\omega_k\Phi_2\omega_k^+}\right) i\omega_k^+.
\end{split}
\end{equation}
The corresponding error is proportional to $\operatorname{Im}([\omega_k\Phi_2\omega_k^+]^{-1})$. Note that for any reasonable value of loss the positive-frequency poles of the $\omega_k^+$ branch in Fig.~\ref{fig:Supp:ColorCodedPoles1}(a) are predominantly real and, thus, the error of approximating the residues by taking the real part is anticipated to be negligible.

Having selected the poles that can act as positive-frequency oscillator poles, we reconstruct their negative-frequency counterparts according to $\omega^-=(-\omega^+)^*$ and $r^-=-(r^+)^*$. Hence, the Lorentzian approximation of the target spectrum for $\Phi_2>0$ and $\omega>0$ takes the form
\begin{equation} \label{eq:supp:LA}
\begin{split}
\tilde{\sigma}_\mathrm{LA}(\omega)=\sum_{k\in\mathbb{P}}
\operatorname{Re}\left(\dfrac{1}{\omega_k\Phi_2\omega_k^+}\right)
\left(\dfrac{i\omega_k^{+}}{\omega-\omega_k^{+}}-\dfrac{(i\omega_k^{+})^*}{\omega-(-\omega_k^{+})^*}\right)
\end{split}
\end{equation}
What remains in $\tilde{\sigma}(\omega)=-i\tan z(\omega)=\tilde{\sigma}_\mathrm{LA}(\omega)+\Delta\tilde{\sigma}(\omega)$ is the error of the Lorentzian approximation and is comprised of four contributions: (i)~the subtraction of the negative frequency counterparts we added in Eq.~\eqref{eq:supp:LA}, (ii)~what is left from taking the real part of the residues, (iii)~the poles omitted from the $\omega_k^+$ branch, and (iv)~the entire $\omega_k^-$ branch.
\begin{equation} \label{eq:supp:LAerror}
\begin{split}
\Delta\tilde{\sigma}(\omega)&=
\sum_{k\in\mathbb{P}}
\operatorname{Re}\left(\dfrac{1}{\omega_k\Phi_2\omega_k^+}\right)
\left(\dfrac{(i\omega_k^{+})^*}{\omega-(-\omega_k^{+})^*}\right)
+\sum_{k\in\mathbb{P}}
i\operatorname{Im}\left(\dfrac{1}{\omega_k\Phi_2\omega_k^+}\right)
\left(\dfrac{i\omega_k^{+}}{\omega-\omega_k^{+}}\right)\\
&+\sum_{k\in\mathbb{Z}\backslash \mathbb{P}}
\dfrac{i}{\omega_k\Phi_2}
\left(\dfrac{1}{\omega-\omega_k^{+}}\right)
-\sum_{k\in\mathbb{Z}}
\dfrac{i}{\omega_k\Phi_2}
\left(\dfrac{1}{\omega-\omega_k^{-}}\right).
\end{split}
\end{equation}
This procedure is schematically depicted in Fig.~\ref{fig:Supp:ColorCodedPoles2}(a)-(c). The three panels depict, respectively, the poles of the conductivity when assuming analytically continued $H'(\omega>0)$ [Fig.~\ref{fig:Supp:ColorCodedPoles2}(a)], the poles of the Lorentzian approximation [Fig.~\ref{fig:Supp:ColorCodedPoles2}(b)], and the poles of the remaining error [Fig.~\ref{fig:Supp:ColorCodedPoles2}(c)]. Note that the error mainly resides in the negative half-plane, far away from the positive-frequency part of the pulse spectrum (apart from the second contribution to the error that is not schematically represented).

%Note that the addition-subtraction of the negative-pole counterparts will produce a smaller error when the negative-pole positions happen to be the same, i.e., when $\omega_a=0$ for $\Phi_2>0$.

The case of negative chirp ($\Phi_2<0$) is entirely analogous [Fig.~\ref{fig:Supp:ColorCodedPoles2}(b)]. In this case, we need to select a subset of the $\omega_k^-$ branch, i.e., the poles that satisfy $0<\operatorname{Re}\omega_k^-<\omega_a$ and are predominantly-real. They will be used for the subsequent Lorentzian approximation (they will play the role of positive-frequency poles of a dampened oscillator) and the corresponding indices are denoted with $k\in\mathbb{P}$ [see Fig.~\ref{fig:Supp:ColorCodedPoles2}(d)].

%A subset of the poles in the  $\omega_k^-$ branch (i) are predominantly real and feature a (ii) positive real part and (iii) negative imaginary part. They will be used for the Lorentzian approximation and the corresponding indices are denoted with $k\in\mathbb{P}$.

For the residues [Eq.~\eqref{eq:supp:OscResidues}] we require $r_k^-$ to be of the form $r_k^-=a(i\omega_k^-)$, with $a\in\mathbb{R}$ and $a>0$. Thus, we approximate the actual residues with
\begin{equation} \label{eq:supp:Residues2}
\begin{split}
r_k^-&=-\dfrac{i}{\omega_k\Phi_2}=
+\dfrac{i\omega_k^-}{\omega_k|\Phi_2|\omega_k^-}
\approx \operatorname{Re}\left(\dfrac{1}{\omega_k|\Phi_2|\omega_k^-}\right) i\omega_k^-.
\end{split}
\end{equation}
Hence, the Lorentzian approximation of the target spectrum for $\Phi_2<0$ and $\omega>0$ takes the form
\begin{equation} \label{eq:supp:LAneg}
\begin{split}
\tilde{\sigma}_\mathrm{LA}(\omega)=\sum_{k\in\mathbb{P}}
\operatorname{Re}\left(\dfrac{-1}{\omega_k\Phi_2\omega_k^+}\right)
\left(\dfrac{i\omega_k^{-}}{\omega-\omega_k^{-}}
-\dfrac{(i\omega_k^{-})^*}{\omega-(-\omega_k^{-})^*}\right)
\end{split}
\end{equation}
and the corresponding error, which comprises (i)~the subtraction of the negative frequency counterparts we added in Eq.~\eqref{eq:supp:LAneg}, (ii)~what is left from taking the real part of the residues, (iii)~the poles omitted from the $\omega_k^-$ branch, and (iv)~the entire $\omega_k^+$ branch, reads
\begin{equation} \label{eq:supp:LAerrorNeg}
\begin{split}
\Delta\tilde{\sigma}(\omega)&=
\sum_{k\in\mathbb{P}}
\operatorname{Re}\left(\dfrac{-1}{\omega_k\Phi_2\omega_k^+}\right)
\left(\dfrac{(i\omega_k^{-})^*}{\omega-(-\omega_k^{-})^*}\right)
+\sum_{k\in\mathbb{P}}
i\operatorname{Im}\left(\dfrac{-1}{\omega_k\Phi_2\omega_k^+}\right)
\left(\dfrac{i\omega_k^{-}}{\omega-\omega_k^{-}}\right)\\
&+\sum_{k\in\mathbb{Z}\backslash \mathbb{P}}
\dfrac{i}{\omega_k\Phi_2}
\left(\dfrac{-1}{\omega-\omega_k^{-}}\right)
-\sum_{k\in\mathbb{Z}}
\dfrac{i}{\omega_k\Phi_2}
\left(\dfrac{1}{\omega-\omega_k^{+}}\right).
\end{split}
\end{equation}

% 
%\section{\label{sec:Supp:TransferFunctionAndError}Error introduced by Lorentzian approximation in conductivity and transfer function}

\begin{figure*}[h]
\includegraphics{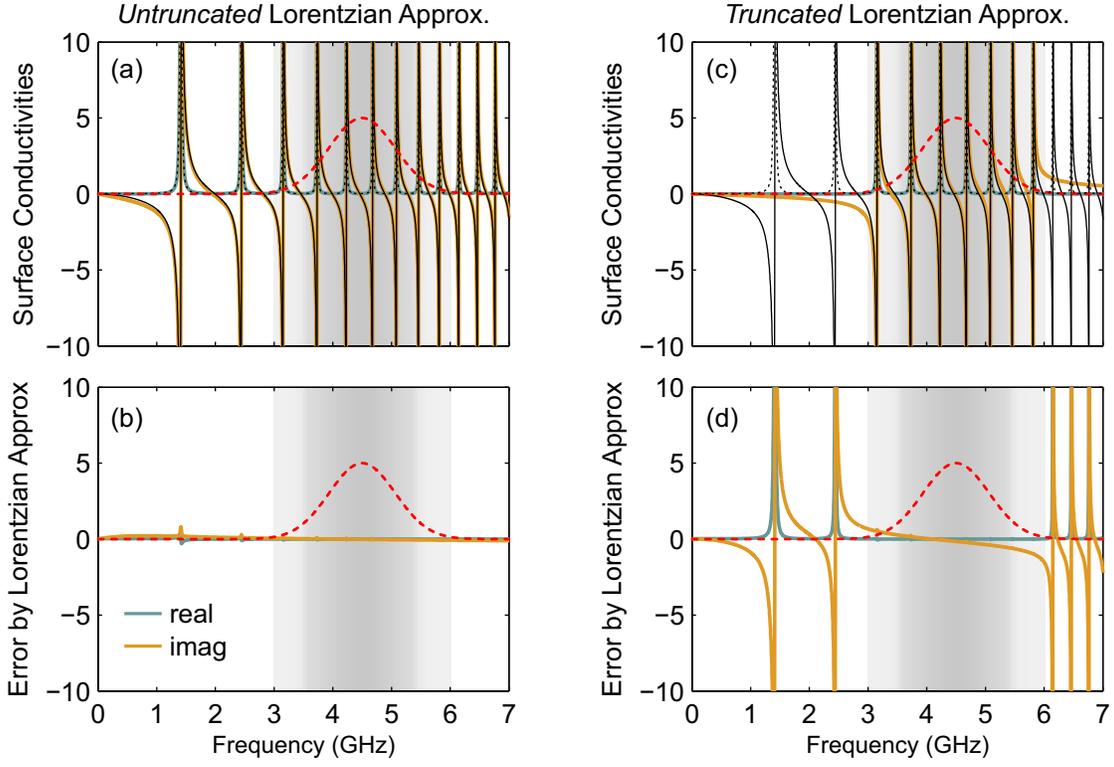}
\caption{\label{fig:Supp:ResFunct1} Quantifying the error introduced by approximating the target conductivity (assuming the discontinuous transfer function $H'(\omega)$) with Lorentzians for the \emph{positive} chirp scenario depicted in Fig.~\ref{fig:ResultPosChirp} of the main text. (a,b)~Untruncated Lorentzian train. (a)~Comparison of conductivity target spectrum and Lorentzian approximation and (b) their difference (error). (c,d)~Truncated Lorentzian train. (c)~Comparison of conductivity target spectrum and Lorentzian approximation and (d) their difference (error). The pulse spectrum is included in all panels with a red dashed line. The error in both cases is benign along the pulse bandwidth. A small contribution with negative slope develops in the truncated case due to the missing resonances. The small kinks that can be discerned in panels (b) and (d) at the positions of the resonances are due to the approximation in the residues (taking their real part).}
\end{figure*}

We can now plot the Lorentzian approximation to the required surface conductivity, as well as the remaining error. The positive chirp case is depicted in Fig.~\ref{fig:Supp:ResFunct1}; the parameters are identical to Fig.~\ref{fig:ResultPosChirp} in the main text. The Lorentzian approximation is depicted in Fig.~\ref{fig:Supp:ResFunct1}(a) and compared with the target spectrum (thin black curves). The error is plotted in Fig.~\ref{fig:Supp:ResFunct1}(b) and is benign across the pulse bandwidth. The small kinks that can be discerned in Fig.~\ref{fig:Supp:ResFunct1}(b) at the positions of the resonances are due to the approximation in the residues (taking their real part),  i.e., the second contribution to the error [cf. Eq.~\eqref{eq:supp:LAerror}].

We next proceed to truncate the infinite Lorentzian train [Fig.~\ref{fig:Supp:ResFunct1}(c),(d)]. This introduces additional error, manifesting as a negative slope in the imaginary part of the conductivity due to the missing resonances; however, if we take care to use a sufficient number of resonances so as to accommodate the pulse bandwidth, the error across the pulse bandwidth is small. It can be even further improved by introducing a counteracting background contribution and/or fine-tuning the positions and strengths of considered resonances.

\begin{figure*}[h]
\includegraphics{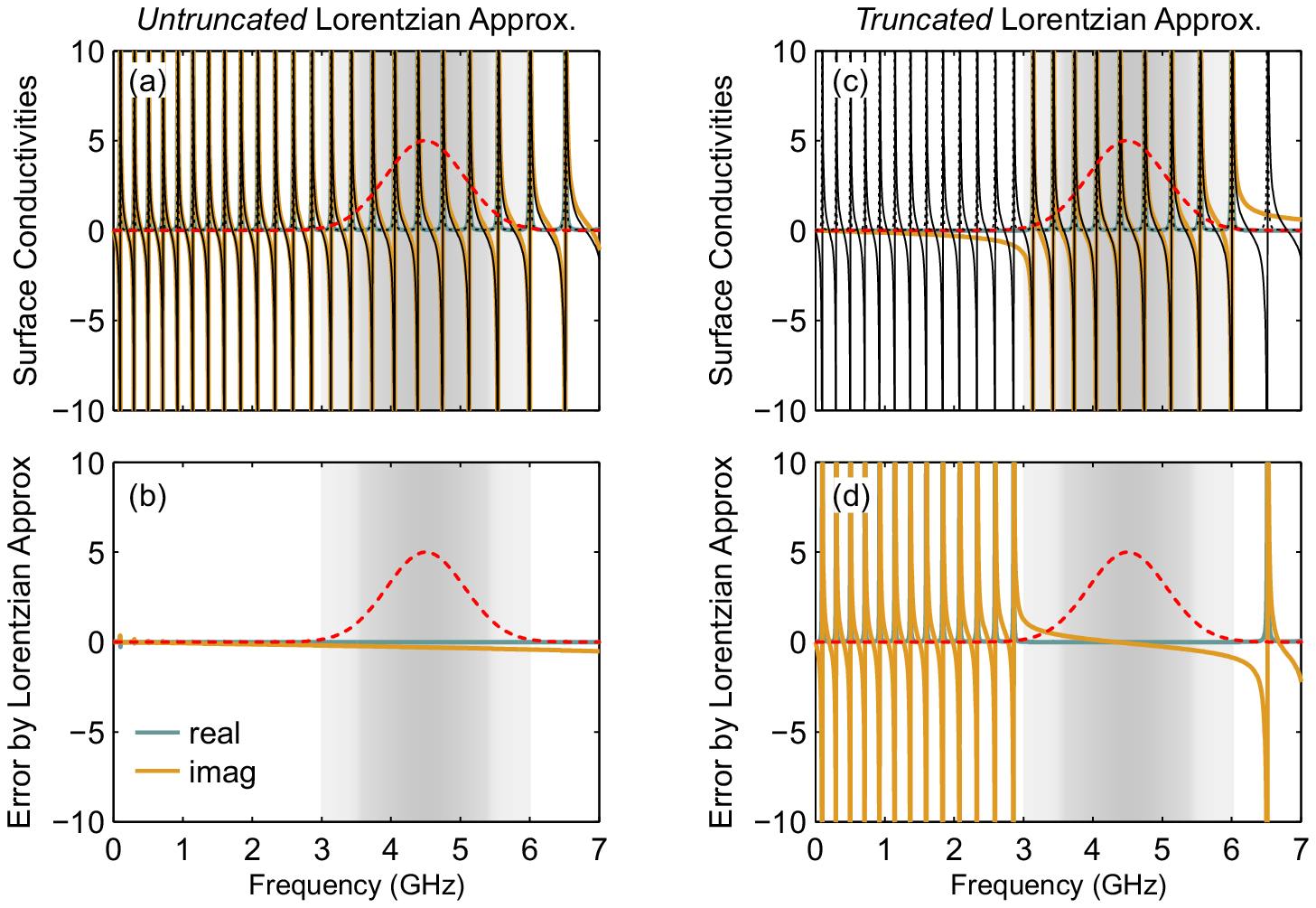}
\caption{\label{fig:Supp:ResFunct2} Quantifying the error introduced by approximating the the target conductivity (assuming the discontinuous transfer function $H'(\omega)$) with Lorentzians for the \emph{negative} chirp scenario depicted in Fig.~\ref{fig:ResultNegChirp} of the main text. (a,b)~Untruncated Lorentzian train. (a)~Comparison of target spectrum  and Lorentzian approximation and (b) their difference (error). (c,d)~Truncated Lorentzian train. (c)~Comparison of target spectrum and Lorentzian approximation and (d)~their difference (error). The pulse spectrum is included in all panels with a red dashed line. The error in both cases is benign along the pulse bandwidth. A small contribution with negative slope develops in the truncated case due to the missing resonances. The small kinks that can be discerned in panels (b) and (d) at the positions of the resonances are due to the approximation in the residues (taking their real part).}
\end{figure*}

Finally, the negative chirp case is depicted in Fig.~\ref{fig:Supp:ResFunct2}; the parameters are identical to Fig.~\ref{fig:ResultNegChirp} in the main text.

\section{\label{sec:Supp:TransfFunctPoles}Analyticity of transfer function in upper complex half-plane}

\begin{figure*}[h]
\includegraphics[width=16cm]{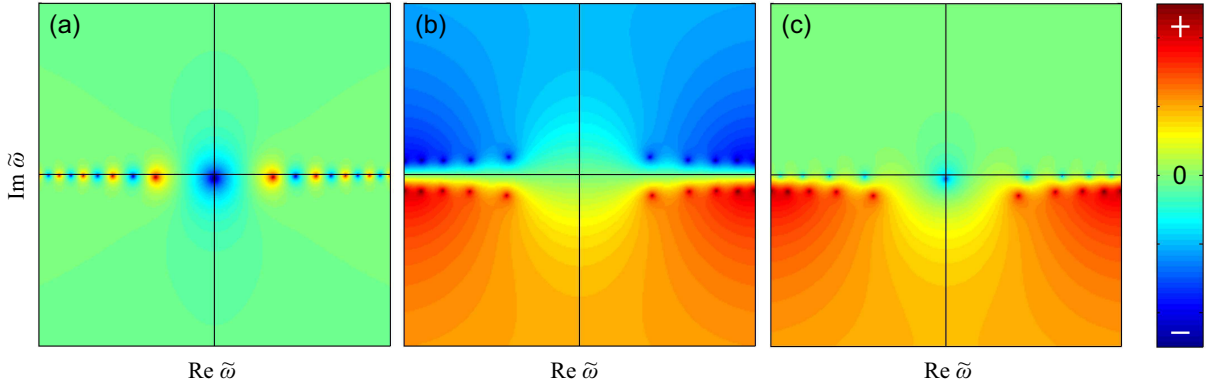}
\caption{\label{fig:Supp:TransfFunctPoles} \green{Visualization of poles/zeros: (a)~The surface conductivity $\tilde{\sigma}_\mathrm{LA}(\omega)$ has both poles and zeros in the lower half-plane; (b)~the transfer function $t=2/(1+\tilde{\sigma})]-1$ has poles in the lower half-plane and zeros in the upper half-plane; (c)~the scattered field $t-1=2/(1+\tilde{\sigma})]-2$ has again all poles and zeros in the lower half-plane. In all cases, we plot the absolute value of the complex function in logarithmic scale. The poles correspond to diverging positive values (dark red) and the zeros to diverging negative values (dark blue).}}
\end{figure*}

\green{
In the main text, we have shown that the proposed response function, i.e., the Lorentzian approximation $\tilde{\sigma}_\mathrm{LA}(\omega)$ given by Eq.~(4), exhibits poles that reside only in the lower complex half-plane. For convenience this is also shown in Fig.~\ref{fig:Supp:TransfFunctPoles}(a) by plotting $|\tilde{\sigma}_\mathrm{LA}(\omega)|$ in logarithmic scale. The poles correspond to diverging positive values and the zeros to diverging negative values. In this Section, we discuss the transfer function, i.e., the transmission or reflection (complex coefficients), which are connected with the surface conductivities via Eqs.~(S18). Let us focus on operation in transmission, meaning that we opt for overlapping conductivities  $\tilde{\sigma}_{se}(\omega)=\tilde{\sigma}_{sm}(\omega)=\tilde{\sigma}(\omega)$. Then, the transmission coefficient given by Eq.~(S18a) becomes
\begin{equation}\label{eq:supp:t1}
t(\omega)=\frac{1-\tilde{\sigma}^2}{1+2\tilde{\sigma}+\tilde{\sigma}^2}%
=\frac{(1-\tilde{\sigma})(1+\tilde{\sigma})}{(1+\tilde{\sigma})^2}%
=\frac{1-\tilde{\sigma}}{1+\tilde{\sigma}}.
\end{equation}
Note that operating in reflection ($\tilde{\sigma}_{se}(\omega)=1/\tilde{\sigma}_{sm}(\omega)=\tilde{\sigma}(\omega))$ results in the very same form, $r(\omega)=-(1-\tilde{\sigma})/(1+\tilde{\sigma})$, and can be treated entirely similarly. 
}

\green{
Next, we can see from the partial fraction decomposition 
\begin{equation}\label{eq:supp:t2}
t(\omega)=\frac{1-\tilde{\sigma}}{1+\tilde{\sigma}}%
=\frac{1}{1+\tilde{\sigma}}-\frac{\tilde{\sigma}}{1+\tilde{\sigma}}%
=\frac{1}{1+\tilde{\sigma}}-\left(1-\frac{1}{1+\tilde{\sigma}}\right)%
=\frac{2}{1+\tilde{\sigma}}-1,
\end{equation}
that the only singularity is found at $\tilde{\sigma}=-1$. Now, we have to find the corresponding frequencies in the complex frequency plane. 
We will show that for a Lorentzian conductivity the real part, $\operatorname{Re} \tilde{\sigma}$, cannot become negative in the upper complex half-plane. This means that the condition $\tilde{\sigma}=-1$ cannot be satisfied in the upper half-plane and the transfer function is analytic there. We focus on a single term of the Lorentzian sum given by Eq.~(4), i.e., 
\begin{equation} \label{eq:LA1}
\tilde{\sigma}_k=\dfrac{i\omega_k}{\omega-\omega_k}
+\dfrac{i\omega_k^*}{\omega+\omega_k^*},
\end{equation}
where we have omitted the real constant prefactor and removed the ``+'' superscript for brevity. It is important to note that for all the terms included in the sum ($k\in\mathbb{P}$) it holds $\operatorname{Re}\omega_k>0$ and $\operatorname{Im}\omega_k<0$. Next, we take the real part of Eq.~\eqref{eq:LA1} and define $\omega=:x+iy$ and $\omega_k=:x_k+iy_k$. We have
\begin{equation} \label{eq:LA2}
\begin{split}
\operatorname{Re}\left\{\dfrac{i\omega_k}{\omega-\omega_k}
+\dfrac{i\omega_k^*}{\omega+\omega_k^*}\right\}&=-\operatorname{Im}\left\{\dfrac{\omega_k}{\omega-\omega_k}
+\dfrac{\omega_k^*}{\omega+\omega_k^*}\right\}\\
&=-\operatorname{Im}\left\{\dfrac{2\omega\operatorname{Re}\omega_k}{\omega^2-|\omega_k|^2-i2\omega\operatorname{Im}\omega_k}\right\}\\
&=-\operatorname{Im}\left\{\dfrac{2x_k(x+iy)}{(x^2-y^2-x_k^2-y_k^2+2yy_k) + i2x(y-y_k)}\right\}\\
&=-\dfrac{2x_k\left[y(x^2-y^2-x_k^2-y_k^2+2yy_k)-2x^2(y-y_k)\right]}
{(x^2-y^2-x_k^2-y_k^2+2yy_k)^2 + 4x^2(y-y_k)^2}.
\end{split}
\end{equation}
The denominator is positive except on the pole $y=y_k \land x=x_k$ (i.e., $\omega=\omega_k$). It is thus positive in the entire upper half-plane, since the pole resides strictly in the lower half-plane. We are, thus, interested in the sign of the numerator. Since $x_k>0$ (for $k\in\mathbb{P}$), we can write
\begin{equation} \label{eq:LA3}
\begin{split}
\operatorname{sgn}\left\{\operatorname{Re}\tilde{\sigma}_k\right\}
&=\operatorname{sgn}\left\{-y[x^2-x_k^2-(y-y_k)^2]+2x^2(y-y_k)\right\}\\
&=\operatorname{sgn}\left\{y[(y-y_k)^2+x_k^2]+x^2(y-2y_k)\right\}.
\end{split}
\end{equation}
The expression in square brackets is positive. Given that $y_k<0$, it can be readily seen that for $y\geq 0$ both terms in the right hand side of Eq.~\eqref{eq:LA3} are positive and it holds $\operatorname{Re}\tilde{\sigma}_k>0$. In other words, we have shown that the condition $\tilde{\sigma}=-1$ cannot be satisfied in the upper half-plane (including the real axis) and the transfer function is analytic there.
}

\green{
To visualize the pole structure of the transfer function given by Eq.~\eqref{eq:supp:t2}, we plot $|[2/(1+\tilde{\sigma})]-1|$ in logarithmic sale [Fig.~\ref{fig:Supp:TransfFunctPoles}(b)]. Clearly, the poles, which are indicated by diverging positive values (dark red), are residing in the lower complex half-plane. 
}

\green{
Regarding the application of the Kramers-Kronig relations between the real and imaginary parts of the transfer function, we note that $t$ does not vanish at infinity (even in the case of a finite sum of Lorentzians). This is not specific to our system: even a sheet of vacuum would lead to the same problem since it would correspond to $t(\omega)=1$. This issue is lifted if we consider the scattered field given by $t-1=[2/(1+\widetilde{\sigma})]-2$ (the total transmitted field is the sum of incident field plus scattered field). In this case, for frequencies extending beyond the finite sum of Lorentzian resonances $\tilde{\sigma}\rightarrow 0$ and, consequently, $t-1\rightarrow 0$. Thus, we conclude that the usual Kramers-Kronig relations apply to the scattered field, $t(\omega)-1$. 
}

\green{
Obviously, since the real and imaginary parts of $t-1$ are related, this will necessarily impose some relation between the corresponding magnitude and phase. In general, this cannot be cast in a simple closed-form relation \cite{Bechhoefer:2011}. However, according to Ref.~\onlinecite{Bechhoefer:2011}, when the function obeys the Kramers-Kronig criteria and moreover does not have zeros in the upper half-plane
%when the function vanishes at infinity and when both poles and zeros are in the lower half-plane, 
the mag-phase relation can take the form of a Hilbert transform pair (Bode relation), similar to the usual Kramers-Kronig relations. In Fig.~\ref{fig:Supp:TransfFunctPoles}(c) we plot the structure of poles/zeroes for $t-1$. It can be seen that both poles and zeros reside in the lower half-plane.
}

\section{\label{sec:Supp:Reflection}Operation in reflection}

\begin{figure*}[h]
\includegraphics{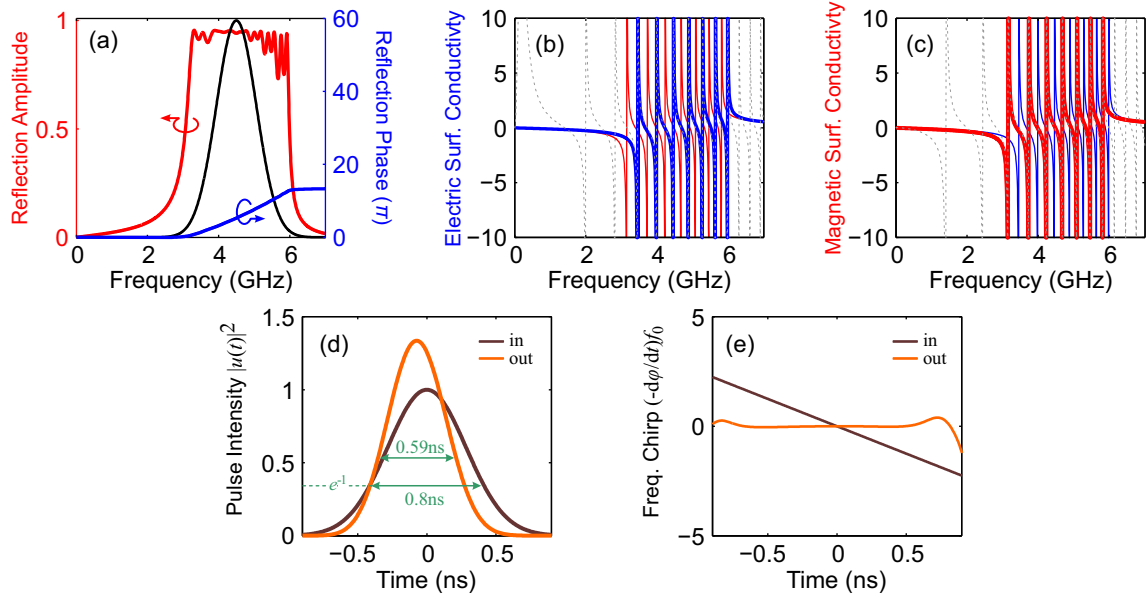}
\caption{\label{fig:Supp:ReflPosChirp} Operation in reflection for the \emph{positive} chirp scenario in Fig.~\ref{fig:ResultPosChirp} of the main text. Truncated recipe resulting in a finite bandwidth. (a)~Reflection amplitude and phase along with  pulse spectrum. Some ripples manifest in the reflection amplitude due to the truncation, primarily at the edges of the band. (b)~\emph{Electric} (dimensionless) surface conductivity (imaginary part). The ideal surface conductivity is included with a dashed line. The magnetic conductivity is also included with a thin solid line. The resonances are interleaved ($\tilde{\sigma}_{sm}=1/\tilde{\sigma}_{se}$). (c)~\emph{Magnetic} (dimensionless) surface conductivity (imaginary part). The ideal surface conductivity is included with a dashed line. and the electric conductivity  with a thin solid line. (d)~Input and output pulse. The output pulse is compressed by approximately $1/\sqrt{2}$, as designed (verified by the durations measured at the $e^{-1}$ intensity points). (e)~Frequency chirp (variation of instantaneous frequency). The output pulse is de-chirped; the residual chirp is negligible along the pulse duration.}
\end{figure*}

In this Section, we present results for operation in reflection to complement the results for operation in transmission presented in the main text (Fig.~\ref{fig:ResultPosChirp}). Figure~\ref{fig:Supp:ReflPosChirp} focuses on the case of \emph{positive} chirp ($\Phi_2>0$). The input pulse is a broadband modulated Gaussian pulse of the form
$u_\mathrm{in}(t)=\exp[-(1+iC)(t-t_0)^2/(2\tau_0^2)]\exp[-i\Omega(t-t_0)]$, with $\Delta\omega=1/\tau_0$ being the transform limited spectral half-width ($e^{-1}$ intensity point) of the pulse spectrum.
%=\exp[-(1+iC)\Delta\omega^2(t-t_0)^2/2]\exp[-i\Omega(t-t_0)]$,
% =e^{-(1+iC)\Delta\omega^2\frac{(t-t_0)^2}{2}}$
%\begin{equation}
% u_\mathrm{in}(t)=e^{-(1+iC)\frac{(t-t_0)^2}{2\tau_0^2}}
% =e^{-(1+iC)\Delta\omega^2\frac{(t-t_0)^2}{2}}
%\end{equation}
The parameters of the example are $\Omega=2\pi\times4.5\cdot10^9$~rad/s, initial chirp $C=-1$, $\Delta\omega=1/\tau_0=2\pi\times0.28\cdot10^9$~rad/s, $\mathcal{A}=0.95$, $\Phi_2=+0.04$~ps$^2$, $\Phi_1=2.26$~ps, $\Phi_0=32$ (equivalently, $\phi_2=0.04$~ps$^2$, $\phi_1=0$, $\phi_0=0$). We focus on the truncated Lorentzian approximation. As a result, the available bandwidth is finite  and  some ripples manifest in the reflection amplitude, primarily at the edges of the band [Fig.~\ref{fig:Supp:ReflPosChirp}(a)]. The electric and magnetic (dimensionless) surface conductivities (imaginary part) are depicted in Fig.~\ref{fig:Supp:ReflPosChirp}(b),(c); they are interleaved ($\tilde{\sigma}_{sm}=1/\tilde{\sigma}_{se}$). The truncation only slightly affects the intended performance: the pulse is compressed by approximately $1/\sqrt{2}$ [Fig.~\ref{fig:Supp:ReflPosChirp}(d)], as designed (verified by the pulse durations measured at the $e^{-1}$ intensity points) and de-chirped [Fig.~\ref{fig:Supp:ReflPosChirp}(e)].

\begin{figure*}[h]
\includegraphics{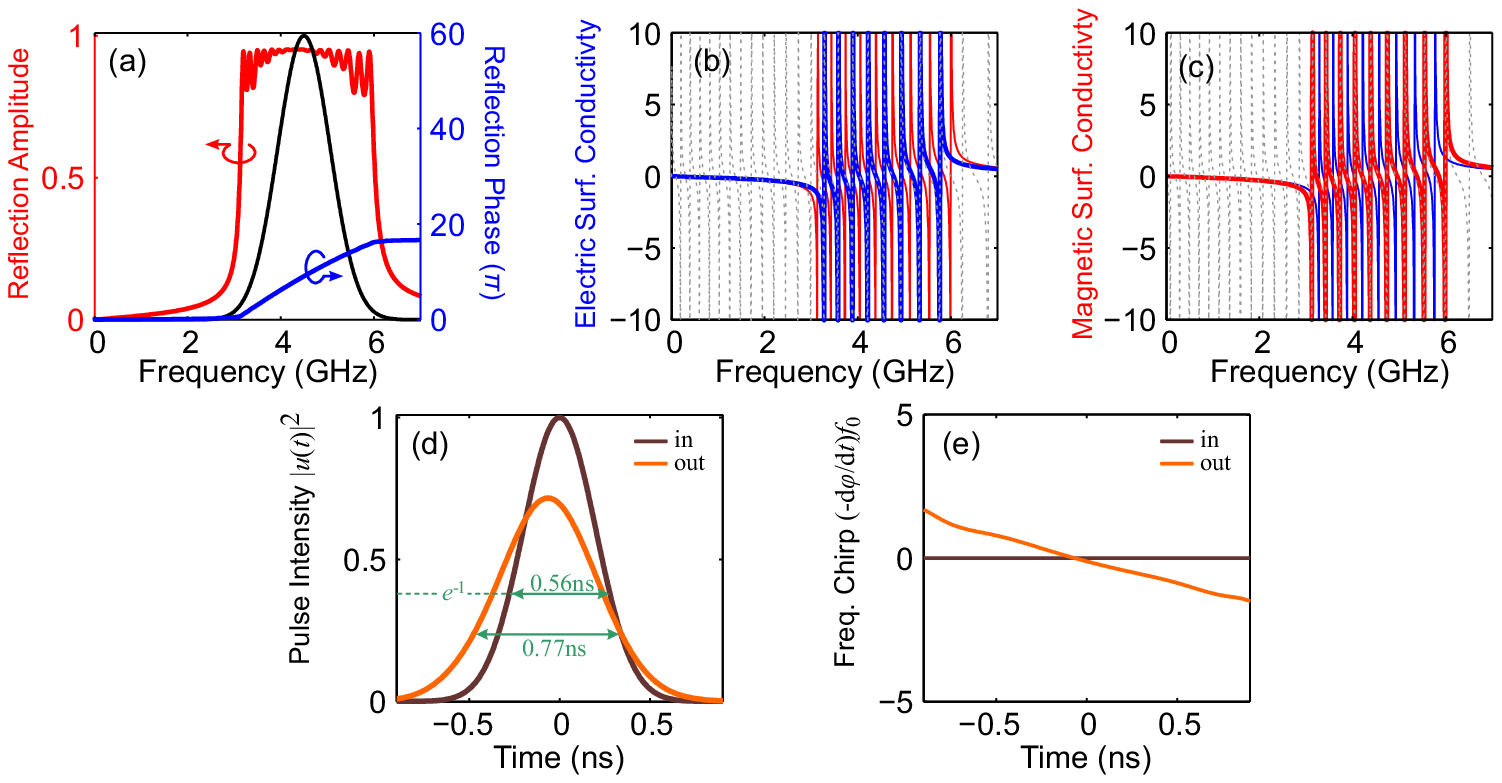}
\caption{\label{fig:Supp:ReflNegChirp} Operation in reflection for the \emph{negative} chirp scenario in Fig.~\ref{fig:ResultNegChirp} of the main text. Truncated recipe resulting in a finite bandwidth. (a)~Reflection amplitude and phase along with  pulse spectrum. Some ripples manifest in the reflection amplitude due to the truncation, primarily at the edges of the band. (b)~\emph{Electric} (dimensionless) surface conductivity (imaginary part). The ideal surface conductivity is included with a dashed line. The magnetic conductivity is also included with a thin solid line. The resonances are interleaved ($\tilde{\sigma}_{sm}=1/\tilde{\sigma}_{se}$). (c)~\emph{Magnetic} (dimensionless) surface conductivity (imaginary part). The ideal surface conductivity is included with a dashed line. The magnetic conductivity is also included with a thin solid line.
(d)~Input and output pulse. The output pulse is broadened by approximately $\sqrt{2}$, as designed (verified by the pulse durations measured at the $e^{-1}$ intensity points). (e)~Frequency chirp (variation of instantaneous frequency. The output pulse acquires a linear chirp along the pulse duration.}
\end{figure*}

Figure~\ref{fig:Supp:ReflNegChirp} shows the case of \emph{negative} chirp ($\Phi_2<0$). The parameters of the example are $\Omega=2\pi\times4.5\cdot10^9$~rad/s, initial chirp $C=0$, $\Delta\omega=1/\tau_0=2\pi\times0.28\cdot10^9$~rad/s, $\mathcal{A}=0.95$, $\Phi_2=-0.04$~ps$^2$, $\Phi_1=2.83$~ps, $\Phi_0=112$ (equivalently, $\phi_2=-0.04$~ps$^2$, $\phi_1=5.1$~ps, $\phi_0=0$). We focus on the truncated Lorentzian approximation. As a result, the available bandwidth is finite  and  some ripples manifest in the reflection amplitude, primarily at the edges of the band [Fig.~\ref{fig:Supp:ReflNegChirp}(a)]. The electric and magnetic (dimensionless) surface conductivities (imaginary part) are depicted in [Fig.~\ref{fig:Supp:ReflNegChirp}(b),(c)]; they are interleaved ($\tilde{\sigma}_{sm}=1/\tilde{\sigma}_{se}$). The truncation only slightly affects the intended performance: the pulse is broadened by approximately $\sqrt{2}$ [Fig.~\ref{fig:Supp:ReflNegChirp}(d)], as designed (verified by the pulse durations measured at the $e^{-1}$ intensity points) and acquires a linear chirp [Fig.~\ref{fig:Supp:ReflNegChirp}(e)].

\clearpage
\twocolumngrid

%\bibliography{AchromaticDualRT}


\end{document}